%% file: A.SN2016aqf.tex
\documentclass[fleqn,usenatbib]{mnras}

\usepackage{newtxtext,newtxmath}
\usepackage[T1]{fontenc}
\usepackage{ae,aecompl}

\usepackage{graphicx}	% Including figure files
\usepackage{subfigure}
\usepackage{amsmath}	% Advanced maths commands
\usepackage{amssymb}	% Extra maths symbols
\usepackage{color}
\usepackage{adjustbox}
\usepackage{natbib}
\usepackage[tableposition=top]{caption}

\usepackage{etoolbox}
\makeatletter
\patchcmd\@combinedblfloats{\box\@outputbox}{\unvbox\@outputbox}{}{%
  \errmessage{\noexpand\@combinedblfloats could not be patched}%
}%
\makeatother

\newcommand{\sn}{SN\,2016aqf}

\newcommand{\nickel}{$^{56}$Ni\,}
\newcommand{\mni}{M$_{\rm Ni}$}
\newcommand{\msun}{M$_{\sun}$}
\newcommand{\halpha}{H$\alpha$}
\newcommand{\eexp}{\ensuremath{\mathrm{E}_\mathrm{exp}}}
\newcommand{\menv}{\ensuremath{\mathrm{M}_\mathrm{env}}}
\newcommand{\rprog}{\ensuremath{\mathrm{R}_\mathrm{prog}}}

\defcitealias{Jerkstrand12}{J12}
\defcitealias{Jerkstrand14}{J14}
\defcitealias{Jerkstrand15a}{J15a}
\defcitealias{Jerkstrand15b}{J15b}
\defcitealias{Jerkstrand18}{J18}

\title[The low-luminosity type II SN\,2016aqf]{The low-luminosity type II SN\,2016aqf: A well-monitored spectral evolution of the Ni/Fe abundance ratio}

\author[M\"uller-Bravo et al.]{
Tom\'as E. M\"uller-Bravo,$^{1}$\thanks{E-mail: t.e.muller-bravo@soton.ac.uk}
Claudia P. Guti\'errez,$^{1}$
Mark Sullivan,$^{1}$
\newauthor
Anders Jerkstrand,$^{2, 3}$
Joseph P. Anderson,$^{4}$
Santiago Gonz\'alez-Gait\'an,$^{5}$
\newauthor
Jesper Sollerman,$^{2}$
Iair Arcavi,$^{6, 7}$ 
Jamison Burke,$^{8, 9}$
Llu\'is Galbany,$^{10}$
\newauthor
Avishay Gal-Yam,$^{11}$
Mariusz Gromadzki,$^{12}$
Daichi Hiramatsu,$^{8, 9}$
\newauthor
Griffin Hosseinzadeh,$^{13}$
D. Andrew Howell,$^{8, 9}$
Cosimo Inserra,$^{14}$
Erki Kankare,$^{15}$
\newauthor
Alexandra Kozyreva,$^{3}$
Curtis McCully,$^{8}$
Matt Nicholl,$^{16, 17}$
Stephen Smartt,$^{18}$
\newauthor
Stefano Valenti,$^{19}$
Dave R. Young$^{18}$
\\
$^{1}$School of Physics and Astronomy, University of Southampton, Southampton, Hampshire, SO17 1BJ, UK\\
$^{2}$The Oskar Klein Centre, Department of Astronomy, Stockholm University, AlbaNova, SE-106 91 Stockholm , Sweden\\
$^{3}$Max-Planck Institut f\"ur Astrophysik, Karl-Schwarzschild Str 1, D-85748 Garching, Germany\\
$^{4}$European Southern Observatory, Alonso de C\'ordova 3107, Casilla 19, Santiago, Chile\\
$^{5}$CENTRA/COSTAR, Instituto Superior T\'ecnico, Universidade de Lisboa, Av. Rovisco Pais 1, P-1049-001 Lisboa, Portugal\\
$^{6}$The School of Physics and Astronomy, Tel Aviv University, Tel Aviv 69978, Israel\\
$^{7}$CIFAR Azrieli Global Scholars program, CIFAR, Toronto, Canada\\
$^{8}$Las Cumbres Observatory, Goleta, California 93117, USA\\
$^{9}$Department of Physics, University of California, Santa Barbara, California 93106, USA\\
$^{10}$Departamento  de  F\'isica  Te\'orica  y  del  Cosmos,  Universidad  deGranada, 18071 Granada, Spain\\
$^{11}$Department of Particle Physics and Astrophysics, Weizmann Institute of Science, Rehovot 76100, Israel\\
$^{12}$Astronomical Observatory, University of Warsaw, Al. Ujazdowskie 4, 00-478 Warszawa, Poland\\
$^{13}$Center for Astrophysics \textbar{} Harvard \& Smithsonian, 60 Garden Street, Cambridge, MA 02138-1516, USA\\
$^{14}$School of Physics \& Astronomy, Cardiff University, Queens Buildings, The Parade, Cardiff, CF24 3AA, UK\\
$^{15}$Department of Physics and Astronomy, University of Turku, Vesilinnantie 5, FI-20014 Turku, Finland\\
$^{16}$Birmingham Institute for Gravitational Wave Astronomy and School of Physics and Astronomy, University of Birmingham,\\Birmingham B15 2TT, UK \\
$^{17}$Institute for Astronomy, University of Edinburgh, Royal Observatory, Blackford Hill, EH9 3HJ, UK \\
$^{18}$Astrophysics Research Centre, School of Mathematics and Physics, Queens University Belfast, Belfast BT7 1NN, UK \\
$^{19}$Department of Physics, University of California, 1 Shields Avenue, Davis, CA 95616-5270, USA\\
}

\date{Accepted XXX. Received YYY; in original form ZZZ}

\pubyear{2020}

\begin{document}
\label{firstpage}
\pagerange{\pageref{firstpage}--\pageref{lastpage}}
\maketitle

\begin{abstract}
Low-luminosity type II supernovae (LL SNe~II) make up the low explosion energy end of core-collapse SNe, but their study and physical understanding remain limited. We present SN\,2016aqf, a LL SN~II with extensive spectral and photometric coverage. We measure a $V$-band peak magnitude of $-14.58$\,mag, a plateau duration of $\sim$100\,days, and an inferred $^{56}$Ni mass of $0.008 \pm 0.002$\,\msun. The peak bolometric luminosity, L$_{\rm bol} \approx 10^{41.4}$\,erg\,s$^{-1}$, and its spectral evolution is typical of other SNe in the class. Using our late-time spectra, we measure the [\ion{O}{i}] $\lambda\lambda6300, 6364$ lines, which we compare against SN II spectral synthesis models to constrain the progenitor zero-age main-sequence mass. We find this to be 12 $\pm$ 3\,\msun. Our extensive late-time spectral coverage of the [\ion{Fe}{ii}] $\lambda7155$ and [\ion{Ni}{ii}] $\lambda7378$ lines permits a measurement of the Ni/Fe abundance ratio, a parameter sensitive to the inner progenitor structure and explosion mechanism dynamics. We measure a constant abundance ratio evolution of $0.081^{+0.009}_{-0.010}$, and argue that the best epochs to measure the ratio are at $\sim$200 -- 300\,days after explosion. We place this measurement in the context of a large sample of SNe II and compare against various physical, light-curve and spectral parameters, in search of trends which might allow indirect ways of constraining this ratio. We do not find correlations predicted by theoretical models; however, this may be the result of the exact choice of parameters and explosion mechanism in the models, the simplicity of them and/or primordial contamination in the measured abundance ratio.
\end{abstract}


%%%%%%%%%%%%%%%%%%%%%%%%%%%%%%%%%%%%%%%%%%%%%%%%%%
%%%%%%%%%%%%%%%%%%%%%%%%%%%%%%%%%%%%%%%%%%%%%%%%%%
%%%%%%%%%%%%%%%%% BODY OF PAPER %%%%%%%%%%%%%%%%%%
%%%%%%%%%%%%%%%%%%%%%%%%%%%%%%%%%%%%%%%%%%%%%%%%%%
%%%%%%%%%%%%%%%%%%%%%%%%%%%%%%%%%%%%%%%%%%%%%%%%%%

\section{Introduction}

Massive stars of $M\gtrsim 8$ \msun\ finish their lives with the collapse of their iron core, which releases great amounts of energy and produces explosions known as core collapse supernovae (CCSNe). These explosions can leave behind compact remnants in the form of neutron stars or black holes, although the exact details of the outcomes are not well understood. Within the different classes of CCSNe, type II SNe (SNe~II), characterised by the presence of hydrogen in their spectra, are the most common \citep{Li11, Shivvers17}. SNe II are a heterogeneous class, with light curves showing different decline rates across a continuum \citep[e.g.,][]{Anderson14} from plateau (SNe~IIP; with a pseudo-constant luminosity for $\sim$\,70 -- 120 days) to linear decliners (SNe~IIL, or fast-declining SNe). The light curves generally show two distinct phases: an optically-thick phase, driven by a combination of the expansion of the ejecta (which pushes the photosphere outwards) and the recombination of hydrogen (which pushes the photosphere inwards), and a later optically-thin phase, powered by the radioactive decay of $^{56}$Co.

SNe~II show a large diversity in luminosities, with peak $V$-band maximum absolute magnitudes ranging from $\sim-13.5$ to $\sim-19$\,mag, and an average of about $-16.7$\,mag \citep[$\sigma$ = 1.01 mag; ][]{Anderson14}. Several low-luminosity SNe~II (LL~SNe~II), generally events with $V\gtrsim-16$\,mag \citep[e.g.,][see also \citealt{GalYam17}; however, note that \citealt{Pastorello12a} proposes an alternative definition]{Kulkarni09, Smartt15a}, have been found in the past decades \citep[e.g.,][]{Turatto98,Pastorello12a,Spiro14,Lisakov18}. 

The prototype of this faint sub-class is SN\,1997D \citep{deMello97, Turatto98}. SN\,1997D displayed a low luminosity and low expansion velocity. However, it was discovered several weeks after peak, with no well-constrained explosion epoch. The first statistical study of this sub-class was that of \citet{Pastorello04}, who found  the class to be characterised by narrow spectral lines (P-Cygni profiles) and low expansion velocities (a few $1000$\,km\,s$^{-1}$ during the late photospheric phase),  suggesting low explosion energies ($E_{\rm exp}\lesssim$ few times $10^{50}$\,erg). Their bolometric luminosity during the recombination ranges between $\sim 10^{41}$ erg s$^{-1}$ and $\sim$\,$10^{42}$ erg s$^{-1}$, with SN\,1999br \citep{Pastorello04} and SN\,2010id \citep{GalYam11} being the faintest SNe~II discovered. They also show lower exponential decay luminosity than the bulk of SNe~II, which reflects their low $^{56}$Ni masses (\mni\ $\lesssim 10^{-2}$ \msun), in agreement with the low explosion energies as expected from the \mni-$E_{\rm exp}$ relation found in different studies \citep[e.g.,][]{Pejcha15b,Kushnir15, Muller17}. \citet{Spiro14} have since expanded the statistical study of LL~SNe~II, adding several objects and finding similar characteristics to those found by \citet{Pastorello04}. While the current sample of nebular spectra of LL~SNe~II is growing, the study of additional events with better cadence and higher signal-to-noise data is essential for understanding their observed diversity.

The progenitors of LL~SNe~II have been shown to be red supergiants (RSGs) with relatively small Zero Age Main Sequence (ZAMS) masses ($M\lesssim15$\,\msun) using archival pre-SN imaging \citep[e.g.][]{Smartt09, Smartt15} and hydrodynamical models \citep[e.g.][]{Dessart13b, Martinez19}. However, other studies have suggested the possibility that their progenitors are more massive RSGs with large amounts of fallback material \citep[e.g.][]{Zampieri03}. Theoretical studies have shown that the nebular [\ion{O}{i}] $\lambda\lambda6300, 6364$ doublet is a good tracer of the progenitor core mass, and, therefore, of the progenitor ZAMS mass \citep[e.g., ][hereafter J12, J14 and J18; and some other studies as well, e.g., \citealt{Lisakov17, Lisakov18}]{Jerkstrand12, Jerkstrand14, Jerkstrand18}, making the late-time spectral evolution extremely important for the study of SN progenitors. Furthermore, nebular nucleosynthesis diagnosis is so far consistent with the lack of massive progenitors above $\sim$20\,\msun\ \citepalias[e.g.][\citealt{Jerkstrand15a, Valenti16}]{Jerkstrand14}.

In addition to the study of the nebular [\ion{O}{i}] $\lambda\lambda6300, 6364$ doublet as progenitor mass estimator, the Ni/Fe abundance ratio, measured from the [\ion{Fe}{ii}] $\lambda7155$ and [\ion{Ni}{ii}] $\lambda7378$ lines, have been shown to be important for the understanding of the inner structure of the progenitor and the explosion mechanism dynamics, as the observed iron-group yields are linked to the temperature, density and neutron excess of the layers that become fuel for the rapid burning process of the explosion \citep[][hereafter J15a, J15b]{Jerkstrand15a, Jerkstrand15b}. However, there are few studies of this ratio, mainly due to the lack of late-time spectra and the absence of these features in the available data in the literature.

In this paper, we study \sn: a well-observed (i.e., excellent spectral and photometric coverage) LL~SN~II, discovered soon after explosion, with $M_V^\mathrm{max} = -14.58$\,mag, a plateau duration of $\sim$\,100 days and a measured \mni\ $= 0.0010$ \msun\ (see Sec.~\ref{subsec:lc} and \ref{subsec:ni_mass}). The nebular spectra show the [\ion{O}{i}] $\lambda\lambda6300, 6364$ doublet. The \ion{He}{i}~$\lambda7065$ emission line is also seen in the spectra of \sn, a line associated to SNe with a low progenitor mass, but not present in every LL~SN~II and not well understood. In addition, \sn\ is one of the few cases where the [\ion{Fe}{ii}] $\lambda7155$ and [\ion{Ni}{ii}] $\lambda7393$ lines (produced by $^{56}$Ni and $^{58}$Ni, respectively) can be seen in the nebular spectra ($\sim$\,150 -- 330 days after the explosion) over $\sim$\,170 days. This extended coverage of the Ni/Fe abundance ratio presents a unique opportunity to study its evolution, and serves as a test for current late-time spectral modelling as well as providing a rich legacy dataset.

This paper is structured as follows: in Sec.~\ref{sec:observations} we describe the observations, data reduction and host galaxy of \sn. In Sec.~\ref{sec:results} we show the light curve, colour and spectral evolution of \sn\ and compare it with other LL~SNe~II. In Sec.~\ref{sec:physical_parameters} we estimate the physical parameters of \sn, while in Sec.~\ref{sec:discussion} we discuss our findings. Finally, our conclusions are in Sec.~\ref{sec:conclusions}. Throughout this paper we assume a flat $\Lambda$CDM cosmology with $H_0=70$\,km\,s$^{-1}$\,Mpc$^{-1}$, $\Omega_\mathrm{M}=0.3$ and $\Omega_{\Lambda}=0.7$, as these values are widely used in the literature \citep[e.g.,][]{Gutierrez18} and the $H_0$ value lies between the value measured from the CMB \citep{Planck15} and local measurements \citep[e.g.,][]{Riess18}.

%%%%%%%%%%%%%%%%%%%%%%%%%%%%%%%%%%%%%%%%%%%%%%%%%%
%%%%%%%%%%%%%%%%%%%%%%%%%%%%%%%%%%%%%%%%%%%%%%%%%%
%%%%%%%%%%%%%%%%%%%%%%%%%%%%%%%%%%%%%%%%%%%%%%%%%%

\section{Observations, reductions and host galaxy}
\subsection{SN Photometry and Spectroscopy}
\label{sec:observations}

\sn\ (ASASSN-16cc) was discovered on 2016 February 26 at 04:33:36 UTC (57444.19 MJD) by the All-Sky Automated Survey for Supernovae\footnote{\url{http://www.astronomy.ohio-state.edu/assassin/index.shtml}} 
\citep[ASAS-SN;][]{Shappee14} at RA = 05h46m23\fs91 and Dec. = $-$52\degr05\arcmin18\farcs9, in NGC\,2101 \citep[][]{Brown16atel} at $z=0.004016$ \citep{Lauberts89}. On 2016 February 27, \sn\ was classified as a SN~II \citep{Hosseinzadeh16Atel,Jha16Atel}. Based on the low luminosity of the host ($M_B=-17.66$\,mag as in \citealt{Gutierrez18}, although see Sec. \ref{sec:host}), we commenced a follow-up campaign with the extended Public ESO Spectroscopic Survey of Transient Objects (ePESSTO) as part of the programme \lq SNe~II in Low-luminosity host galaxies\rq. 

The final pre-explosion non-detection in the $V$-band, reported three days before the date of classification by ASAS-SN (57442 MJD), has a limiting magnitude $\sim16.7$\,mag, which does not give a stringent constraint on the explosion epoch. Previous non-detections have similar limiting magnitudes.
Hence, we decided to estimate the explosion epoch using the spectral matching technique \citep[e.g.,][]{Anderson14, Gutierrez17a}. We used \textsc{gelato}\footnote{\url{https://gelato.tng.iac.es/gelato/}} \citep{Harutyunyan08} to find good spectral matches to the highest resolution spectrum of \sn, as it is also one of the first spectra taken (57446 MJD, see below). From the best matching templates, we calculated a mean epoch of the spectrum of $\sim$6 days after explosion and a mean error added with the standard deviation of the explosion epochs in quadrature of $\sim$ 4 days. This gives an explosion epoch of MJD $57440.19\pm4$ (slightly different to the estimated epoch in \citealt{Gutierrez18} as they used the non-detection)

\begin{figure}
    \includegraphics[width=\columnwidth]{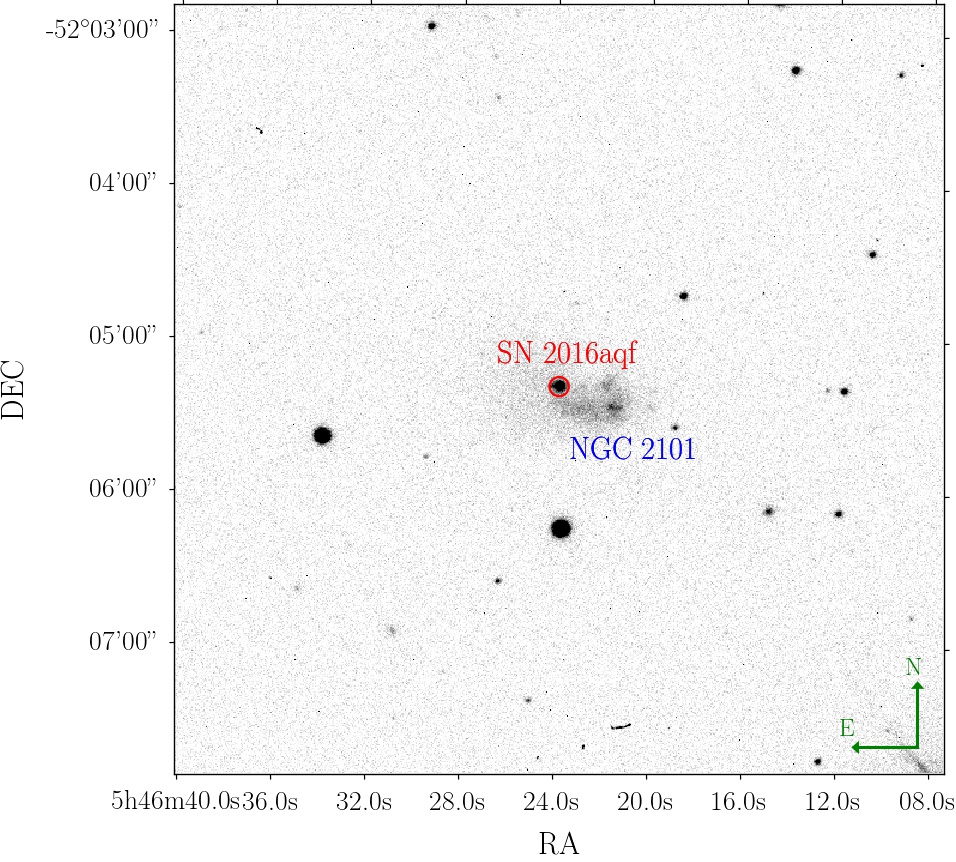}
    \caption{$r$-band image of NGC 2101 with \sn\ marked. Data from the 1.0-m Las Cumbres Observatory telescopes (MJD = 57514, 74 days after explosion).}
    \label{fig:host_image}
\end{figure}

Optical $BVgri$ imaging of \sn\ was obtained with the 1.0-m telescope network of the  Las Cumbres Observatory \citep[LCO;][]{Brown13} as part of both ePESSTO and the \lq Las Cumbres Observatory SN Key Project\rq, with data taken from 8 to 311\,d after explosion. All photometric data were reduced following the prescriptions described by \citet{Firth15}. This pipeline subtracts a deep reference image constructed using data obtained in the $BVgri$ bands three years after the first detection of \sn\ to remove the host-galaxy light using a point-spread-function (PSF) matching routine. SN photometry is then measured from the difference images using a PSF-fitting technique. Fig.~\ref{fig:host_image} shows the SN position within the host galaxy. The photometry of SN\,2016aqf is presented in Table~\ref{tab:photometry}.% in Appendix \ref{app:phot}.

\begin{center}
\input{photometry.tex}
\end{center}

Spectroscopic observations were obtained with the ESO Faint Object Spectrograph and Camera version 2  \citep[EFOSC2;][]{Buzzoni84} at the 3.58-m ESO New Technology Telescope (NTT), the FLOYDS spectrograph \citep{Brown13} on the Faulkes Telescope South (FTS), and the Robert Stobie Spectrograph (RSS; \citealt{Burgh03, Kobulnicky03}) at the Southern African Large Telescope (SALT). FLOYDS spectra were taken as part of the \lq Las Cumbres Observatory SN Key Project\rq. The observations include phases from 2 to 348\,d after explosion. EFOSC2 spectra, obtained with grism \#13, cover 3500--9300 \AA\ at a 21.2\,\AA\ resolution, the FLOYDS spectra 
%has a prism cross-disperser, which images first and 
%second orders onto the chip, resulting in a very
have wavelength coverage of $\sim$\,3200 -- 10000\,\AA\ with a resolution of $\sim18$\,\AA, and the RSS spectrum \citep{Jha16Atel} covers 3600--9200\,\AA\ at $\sim7$\,\AA\ resolution. The data reduction of the EFOSC2 spectra was performed using the PESSTO pipeline\footnote{\url{https://github.com/svalenti/pessto}} \citep{Smartt15a}, while the FLOYDS data were reduced using the \textsc{pyraf}-based \textsc{floydsspec} pipeline\footnote{\url{https://github.com/svalenti/FLOYDS\_pipeline}} \citep{Valenti14}. All spectra are available via the WISeREP\footnote{\url{https://wiserep.weizmann.ac.il/}} repository \citep{Yaron12}. Spectral information is summarised in Table~\ref{tab:spectra_info}.% in Appendix \ref{app:spec_tables}.

\begin{center}
\input{spectra_info.tex}
\end{center}

%%%%%%%%%%%%%%%%%%%%%%%%%%%%%%%%%%%%%%%%%%%%%%%%%%%%%
\subsection{Host Galaxy}
\label{sec:host}

Photometry of NGC2101 was obtained with the LCO 1.0-m telescope network, and spectroscopy with VLT/FORS2, around three years after the SN explosion (2019 February 6 at 04:38:48 UTC). We estimated a galaxy distance of $\mu = 30.16 \pm 0.27$\,mag (see Sec.~\ref{subsec:lc}), consistent with the Tully-Fisher value of $\mu = 30.61 \pm 0.80$\,mag, as reported in the NASA/IPAC Extragalactic Database\footnote{\url{http://ned.ipac.caltech.edu/}} (NED). Adopting the distance estimated in this work, the galaxy has $M_B$ = $-17.22\pm0.34$\,mag, which is consistent with the value reported in \citet[][$-17.66$\,mag]{Gutierrez18} given the large uncertainties from the reported distance. We use the total apparent corrected $B$-magnitude, with the total B-magnitude error as reported in HyperLEDA, using error propagation. The radial velocity corrected for Local Group infall onto Virgo is $883 \pm 3$ km s$^{-1}$ \citep{Theureau98, Terry02}, as reported in HyperLEDA, a value which we use to estimate the corrected redshift of \sn. 

From the spectrum of the \ion{H}{ii} region at the position of the SN, we measure the emission line fluxes of H\,$\alpha$, H\,$\beta$, [\ion{N}{ii}] and [\ion{O}{iii}]. We estimate the star-formation rate (SFR) from the H$\alpha$ line as SFR = $2.3 \pm 0.6 \times 10^{-1}$\,\msun\,yr$^{-1}$ using the calibration from \citet{Kennicutt12}, where the uncertainty is driven by the uncertainty in the distance.
Using the calibration of \citet{Marino13}, we then estimate a gas-phase metallicity of (12 + log(O/H))$_{\rm O3N2} = 8.144 \pm 0.025$ dex and (12 + log(O/H))$_{\rm N2} = 8.134 \pm 0.042$ dex, i.e., below the solar value of 8.69 dex \citep[][]{Asplund09}. This is low compared to many other SN~II host galaxies, but not uncommon \citep[e.g.,][]{Anderson16}. However, the metallicity does not follow the relation found with the \ion{Fe}{ii}~$\lambda5018$ pEW \citep[e.g.,][]{Dessart14b, Anderson16, Gutierrez18}. This may be caused by the lower temperatures in LL~SNe~II which causes the earlier appearance of the \ion{Fe}{ii} lines in these objects \citep{Gutierrez17a}.

%%%%%%%%%%%%%%%%%%%%%%%%%%%%%%%%%%%%%%%%%%%%%%%%%%
%%%%%%%%%%%%%%%%%%%%%%%%%%%%%%%%%%%%%%%%%%%%%%%%%%

\section{Results and analysis}
\label{sec:results}

%%%%%%%%%%%%%%%%%%%%%%%%%%%%%%%%%%%%%%%%%%%%%%%%%%

\subsection{Extinction corrections}
\label{subsec:reddening}

 We adopt a Milky Way extinction value of $E(B-V)_{\mathrm{MW}}=0.047$\,mag, and correct our photometry using the prescription of \citet{Schlafly11} and the \citet{Cardelli89} reddening law with $R_V=3.1$.
% , using the \textsc{sfdmap}\footnote{\url{https://github.com/kbarbary/sfdmap}} and \textsc{extinction}\footnote{\url{https://github.com/kbarbary/extinction}} Python packages.
To estimate the host galaxy extinction, we investigated the equivalent-width (EW) of the \ion{Na}{i}\,D ($\lambda\lambda5889, 5895$) absorption, a well-known tracer of gas, metals and dust \citep[e.g.,][]{Richmond94, Munari97, Turatto03, Poznanski12}. We note that these relations tend to have large uncertainties. 

The spectrum at +6\,d is the only one that seems to shows \ion{Na}{i}\,D absorption lines from the MW and the host galaxy. We used the relations for one line (D$_1$) and two lines (D$_1$+D$_2$) from \citet{Poznanski12}, obtaining upper limits of $E(B-V) \lesssim 0.028 \pm 0.011$\,mag and $E(B-V) \lesssim 0.032 \pm 0.006$\,mag, respectively. This gives a weighted average value of $E(B-V) \lesssim 0.031$\,mag. Given this very small level of extinction (and its uncertainty), we choose not to make an extinction correction to the SN data. We do not use other methods to estimate this value as they rely on the SN colour; \citet{deJaeger18} showed that the majority of colour dispersion of SNe~II is intrinsic to the SN.

\subsection{Light curve and distance}
\label{subsec:lc}

\begin{figure}
	\includegraphics[width=1.0\columnwidth]{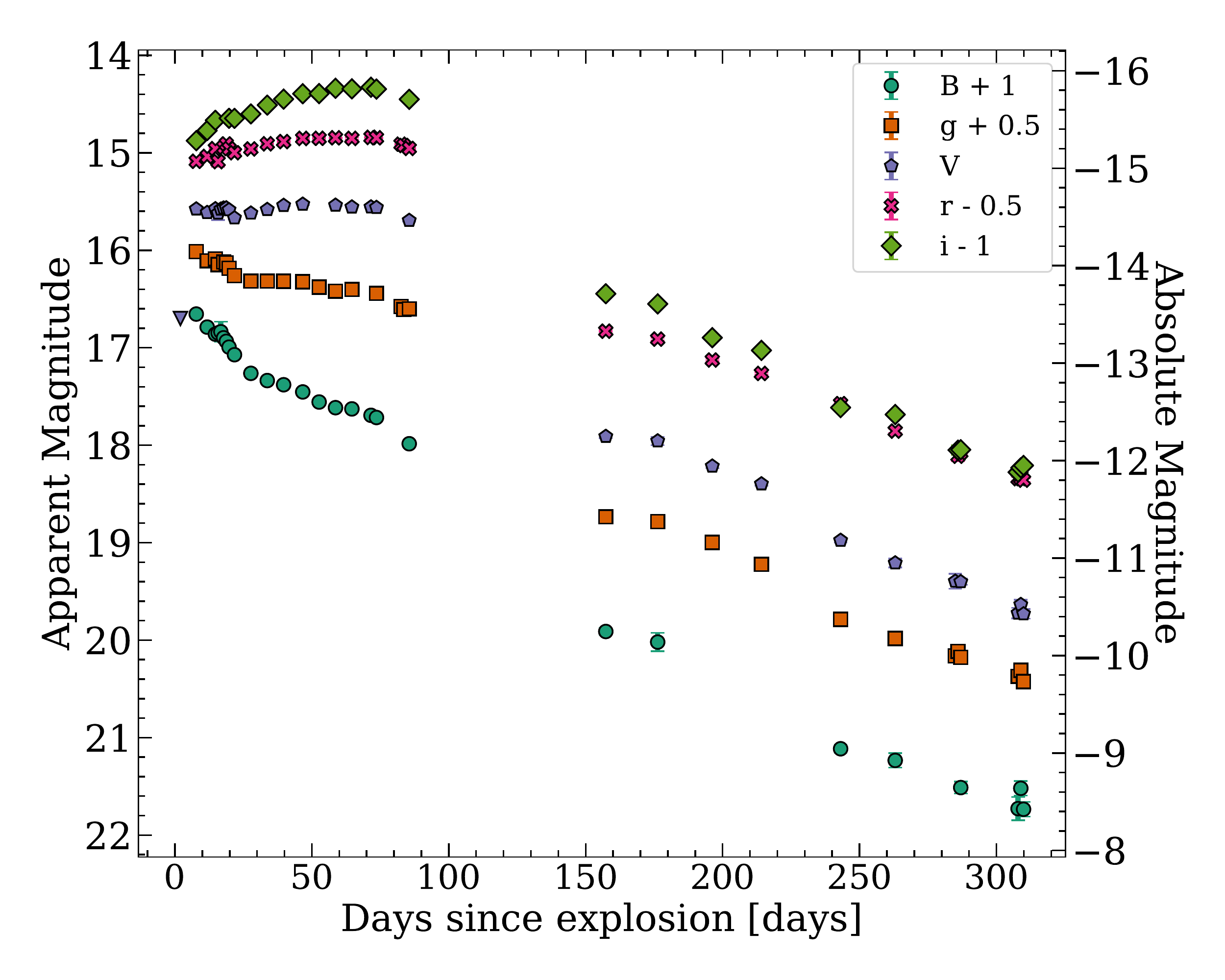}
    \caption{\sn\ $BVgri$-band photometry from +8 to + 311\,d. $BV$ bands are in Vega magnitude system, while $gri$ bands are in AB magnitude system. The last non-detection in $V$ band is also shown (inverted triangle). The SN was not visible around the transition from the optically-thick to the optically-thin phase. Offsets have been applied to the photometry for visualisation purposes. As in all figures in this paper, the photometry is corrected for MW extinction but not host extinction, and the data are in the rest-frame.}
    \label{fig:photometry}
\end{figure}

The $BVgri$-band light curves of SN\,2016aqf (Fig.~\ref{fig:photometry}) cover 8 to 311\,d after explosion (all phases in this paper are relative to the estimated explosion epoch). As the host galaxy is not in the Hubble flow, we estimated the distance to \sn\ using the Standardized Candle Method \citep{Hamuy02L}, which relates the velocity of the ejecta of a SN II to its luminosity during the plateau, and the relation of \citet[][equation 17]{Kasen09} for a redshift-independent distance estimate. We calculate the distance modulus $\mu=30.16\pm0.27$\,mag ($10.8\pm1.4$\,Mpc), which gives $M_V^\mathrm{max}=-14.58$\,mag and a mid-plateau $V$-band luminosity of $-14.63$\,mag (note the plateau luminosity is slightly brighter; $M_V^{\rm max}$ represents the maximum luminosity from the peak closest to the bolometric peak). We estimated $M_V^{\rm max}$ from the first epoch of photometry given that the last non-detection helps to obtain a good constrain.

During the recombination phase, the SN shows an increase in the $Vri$-bands luminosity, probably due to its low temperature which shifts the peak luminosity from the ultraviolet (UV) to redder bands more rapidly compared to normal SNe~II. The gap in observations between 80 and 150 days was caused by the SN going behind the sun, and coincides with the SN transitioning from the optically-thick to the optically-thin phase. The $V$-band decreases by $\sim2$\,mag across the gap in the light curve, and is an estimate of the decrease caused by the transition from plateau to nebular phase, smaller than other LL~SNe~II \citep[$\sim3$--$5$\,mag; e.g.,][]{Spiro14}. We measured the decline rate in the $V$-band at early epochs ($t\lesssim20$ days; $s_1$), in the plateau ($s_2$), and in the exponential decay tail ($s_3$)
%, together with M$_{\rm end}$, NO PUEDO MEDIR ESTO SIN M_tail
as defined in \citet[][see section \ref{subsec:physical_parameters} for the $t_{\rm pt}$ used]{Anderson14}, obtaining $s1=0.65_{-0.12}^{+0.13}$\,mag\,100\,d$^{-1}$, $s2=-0.08_{-0.01}^{+0.01}$ mag 100 d$^{-1}$ and $s3=1.22_{-0.02}^{+0.02}$ mag 100 d$^{-1}$. $M_{\rm tail}$ was not measured as the early decline of the exponential decay tail was not observed.

%%%%%%%%%%%%%%%%%%%%%%%%%%%%%%%%%%%%%%%%%%%%%%%%%%

\subsection{Colour Evolution}
\label{subsec:colour_evolution}

\begin{figure}
\includegraphics[width=\columnwidth]{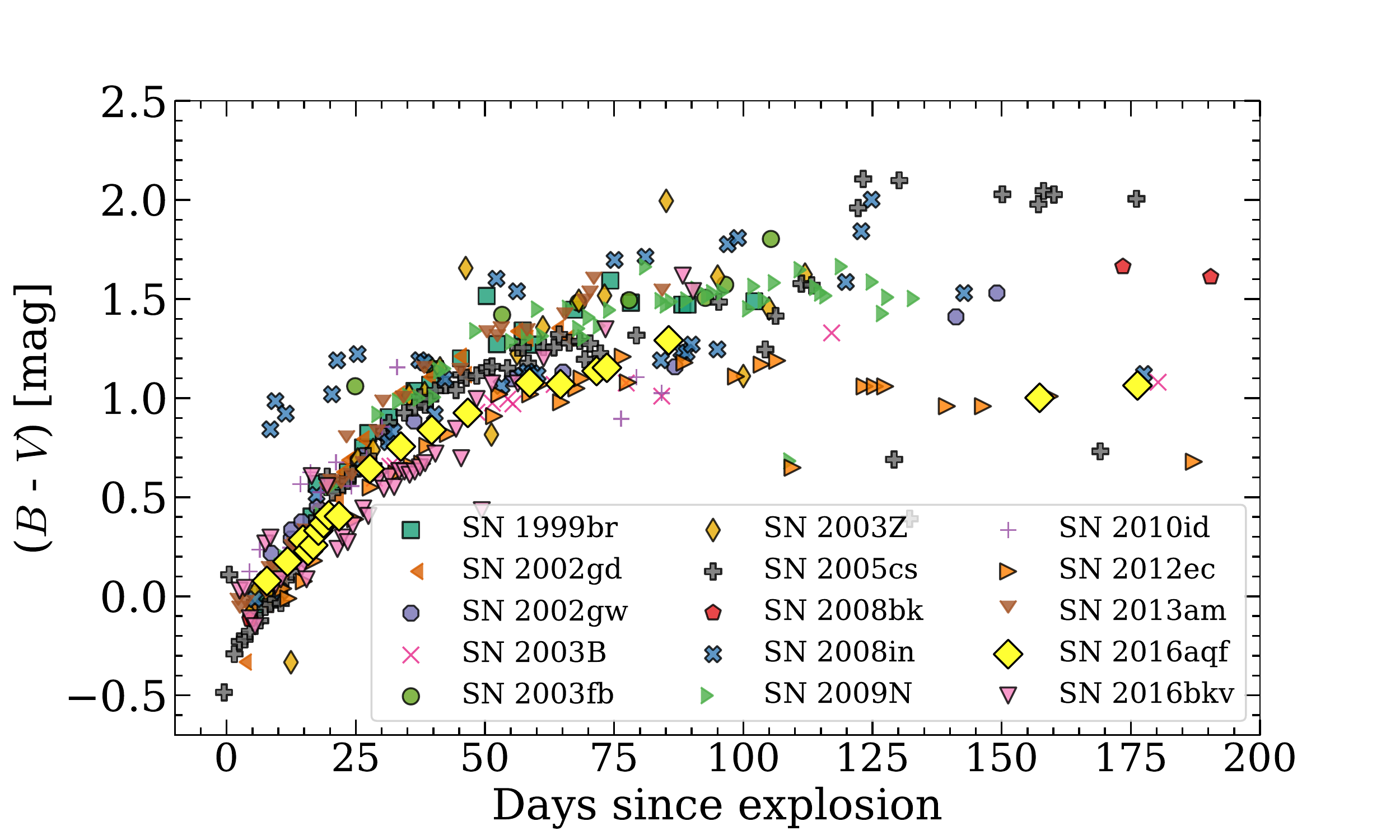}
\caption{($B$-$V$) colour evolution of \sn\ compared to our sample of LL~SNe~II. All data is corrected for MW and host galaxy extinction (except for those SNe with host extinction values reported as upper limits in Table \ref{tab:sn_sample}.) Notice that the dispersion generally increases with time. Uncertainties are not shown as they are relatively small in general.}
\label{fig:colour_evolution}
\end{figure}

In Fig.~\ref{fig:colour_evolution} we show the ($B-V$) colour curve (corrected for MW extinction) of \sn\ during the first 200 days. At the beginning of the observations ($8$\,d) it has a colour close to $0$\,mag, which slowly increases to around $1.0$ mag at $\sim+50$\,d and $\sim1.3$\,mag before the gap in coverage.

For comparison, we form a sample of other LL~SNe~II from the literature with good data coverage and similar properties to our object: SN\,1999br \citep{Hamuy03, Pastorello04, Gutierrez17a}, SN\,2002gd, SN\,2002gw , SN\,2003B, SN\,2003fb, SN\,2003Z, SN\,2004fx, SN\,2005cs, SN\,2008bk, SN\,2008in, SN\,2009N, SN\,2010id, SN\,2013am and SN\,2016bkv. These SNe and their references are in Table~\ref{tab:sn_sample}. In addition we include SN\,2012ec \citep{Maund13}, a non-LL~SN~II, as a reference as it has a well-measured Ni/Fe abundance ratio, used in our later analysis. For this comparison sample, we use photometry and spectra obtained from the \lq Open Supernova Catalog\rq\ \citep{Guillochon17} and WISeREP \citep{Yaron12}. Note that we only used epochs with both $B$ and $V$ photometry to calculate colour, without applying interpolations. The photometry of this sample is corrected for MW extinction (see Sec. \ref{subsec:reddening}), and host galaxy extinction, using the values from the references in Table~\ref{tab:sn_sample}. However, we do not correct for host galaxy extinction when the reported value is an upper limit (this does not represent a problem given the relatively small extinction values, A$_V$ < 0.1 mag).

\begin{center}
\input{sn_sample.tex}
\end{center}

The ($B-V$) evolution of \sn\ is in general flatter than the bulk of our sample, showing similar colours at early epochs ($t\lesssim15$ days), but becoming slightly bluer at later epochs ($t\gtrsim25$ days), similar to SN\,2012ec. After $\sim$\,100\,d the dispersion in the colour evolution of our sample starts increasing, probably due to the faintness of these objects.

%%%%%%%%%%%%%%%%%%%%%%%%%%%%%%%%%%%%%%%%%%%%%%%%%%

\subsection{Bolometric light curve}
\label{subsec:lbol}

\begin{figure*}
	\includegraphics[width=0.9\textwidth]{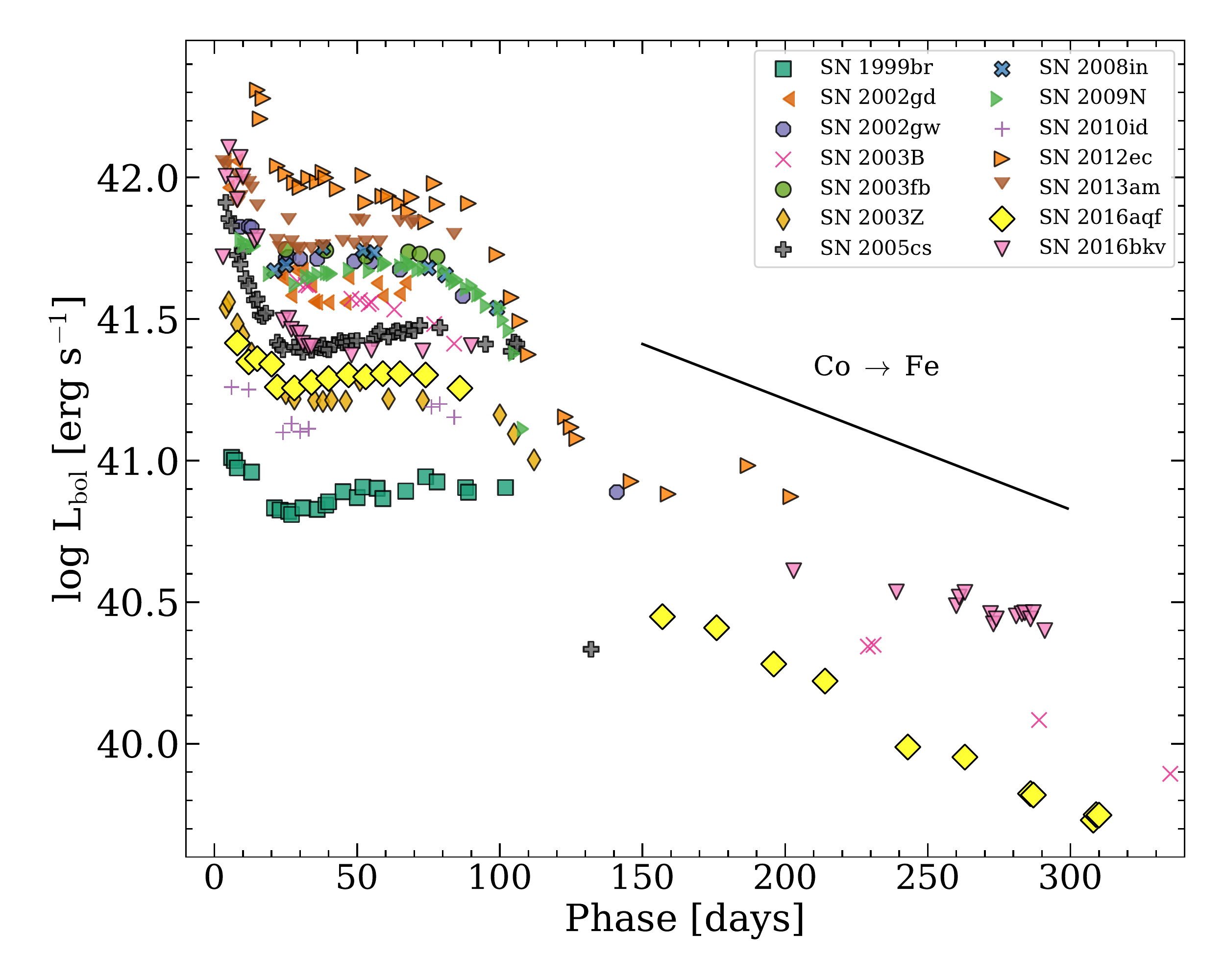}
    \caption{Bolometric light curve of \sn\ compared to our LL~SNe~II sample. The light curves were obtained by using bolometric corrections (see Section~\ref{subsec:lbol} for details). All data is corrected for MW and host galaxy extinction (except for those with values reported as upper limits, see Table \ref{tab:sn_sample}). The $^{56}$Co $\rightarrow$ $^{56}$Fe decay line is shown for comparison. Uncertainties are not shown for visualisation purposes}
    \label{fig:lbol}
\end{figure*}

We estimated the bolometric light curve of \sn\ by applying the bolometric correction from \citet{Lyman14} (assuming a cooling phase of 20 days). We use the ($g-i$) colour as it shows the smallest dispersion. Most SNe in our LL~SN~II sample have only $BVRI$ data, so, to be consistent, we calculated their bolometric light curves (correcting for MW extinction only) by applying the relation from \citet{Lyman14} as well, but with ($B-I$) colour as it has the smallest dispersion within the available bands, using the distances from Table \ref{tab:sn_sample}. Only epochs with simultaneous $B$ and $I$ bands (or $g$ and $i$ for \sn) were used. The light curves are shown in Fig.~\ref{fig:lbol} (SN\,2008bk is not shown as it does not have epochs with simultaneous $B$ and $I$ coverage). Unfortunately, as the relations from \citet{Lyman14} only work in a given colour range, we can not estimate the bolometric light curve during the nebular phase of some of the SNe.

The luminosity of \sn\ at peak is $L_{\rm bol} \approx 10^{41.4}$\,erg\,s$^{-1}$, estimated from the first epoch with photometry. The luminosity of \sn\ during the cooling phase generally decreases less steeply than other LL~SNe~II. During the plateau phase, the luminosity falls to $L_{\rm bol} \approx 10^{41.3}$\,erg\,s$^{-1}$, placing it in the mid-luminosity range of our sample (between SN\,2005cs and SN\,2002gd). After the gap, the SN has a luminosity of $L_{\rm bol}$ $\approx$ $10^{40.5}$\,erg\,s$^{-1}$, dropping to $L_{\rm bol}$ $\approx$ $10^{39.7}$\,erg\,s$^{-1}$ at +300\,d. The exponential tail is steeper than $^{56}$Co decay \citep[0.98\,mag per 100 days][]{Woosley89}, although shallower than the decay in the $V$-band, presumably due to $\gamma$-ray leakage.

%%%%%%%%%%%%%%%%%%%%%%%%%%%%%%%%%%%%%%%%%%%%%%%%%%

\subsection{Early spectral evolution}
\label{subsec:early_spec}

\begin{figure*}
	\includegraphics[width=\textwidth]{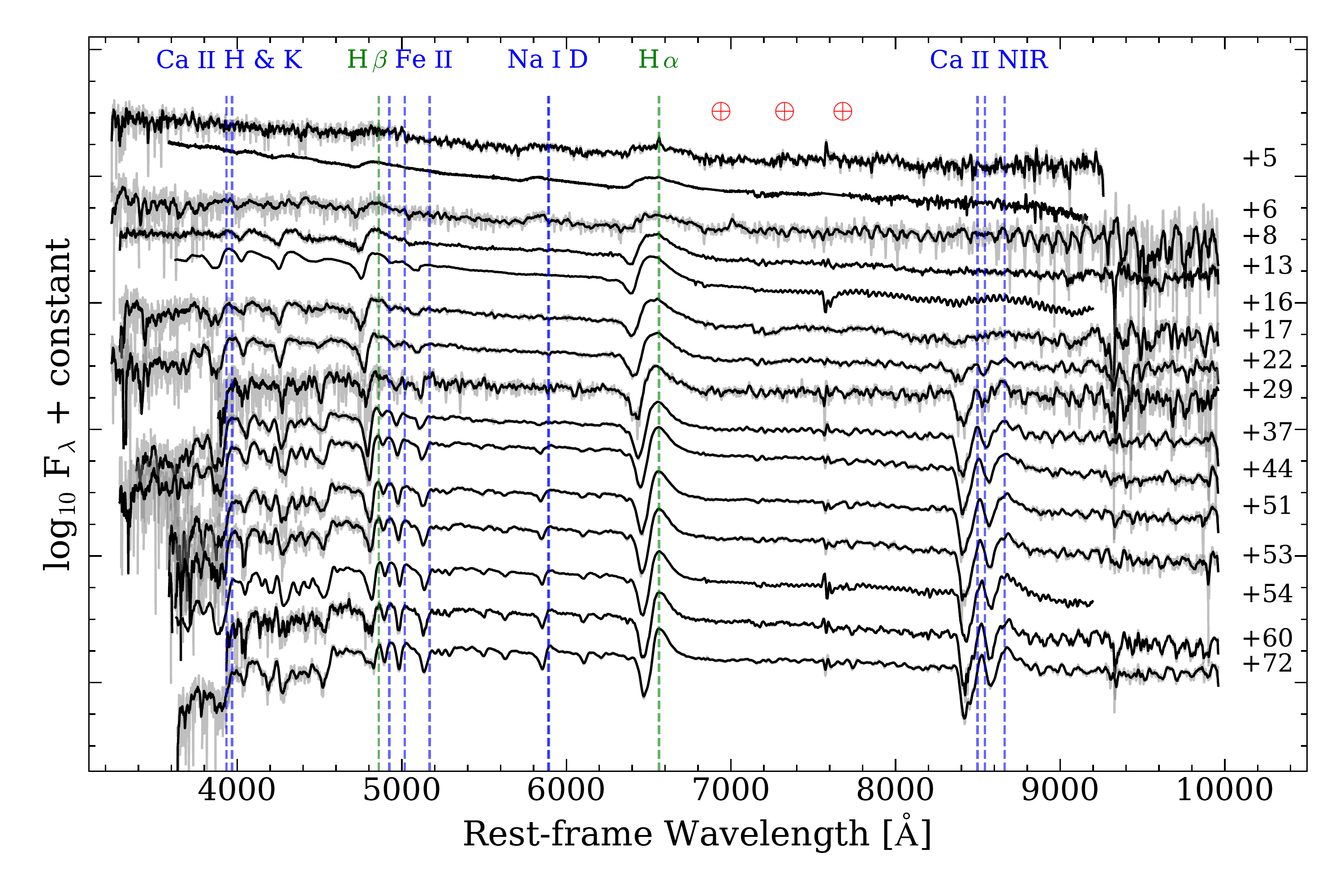}
    \caption{\sn\ photospheric phase spectra. \ion{Ca}{ii} H\&K, H$\beta$, \ion{Fe}{ii}~$\lambda\lambda\lambda4924, 5018, 5169$, \ion{Na}{i}\,D, H$\alpha$ and \ion{Ca}{ii}\,NIR lines are marked. Green vertical lines denote single lines, while blue denotes doublets or triplets. Telluric lines are shown by red circles with crosses. In some cases, the binned spectra (black line) are over-plotted on the original spectra (grey) for visualisation. Spectra corrected for MW extinction.}
    \label{fig:photospheric_spectra}
\end{figure*}

The spectra of \sn\ have narrower lines than spectra of normal SNe~II, suggesting low expansion velocities and low explosion energies. Spectra obtained during the optically-thick phase are shown in Fig.~\ref{fig:photospheric_spectra}. During the first two weeks, the evolution is mainly dominated by a blue continuum and Balmer lines, showing P-Cygni profiles of H$\alpha$ and H$\beta$.
%\ion{He}{i}~$\lambda5876$. 
\ion{Fe}{ii}~$\lambda4924$, $\lambda5018$, $\lambda5169$ and \ion{Ca}{ii}~$\lambda\lambda\lambda8498, 8542, 8662$ then appear, becoming prominent at later epochs. The \ion{Na}{i}\,D appears at around one month. \ion{Sc}{ii}/\ion{Fe}{ii}~$\lambda5531$, \ion{Sc}{ii}~$\lambda5663$, $\lambda6247$ and \ion{Ba}{ii}~$\lambda6142$ appear at around +50\,d. %Although \ion{Sc}{ii}~$\lambda6247$ and \ion{Ba}{ii}~$\lambda6142$ are common lines in LL~SNe~II at this epoch, they are weaker in \sn\ \red{(compare, e.g., with SN~2005cs in Fig.~\ref{fig:photospheric_spectra})}.
\ion{O}{i} $\lambda7774$ is weakly present after one month.

\begin{figure*}
    \centering
    \subfigure{\includegraphics[scale=0.28]{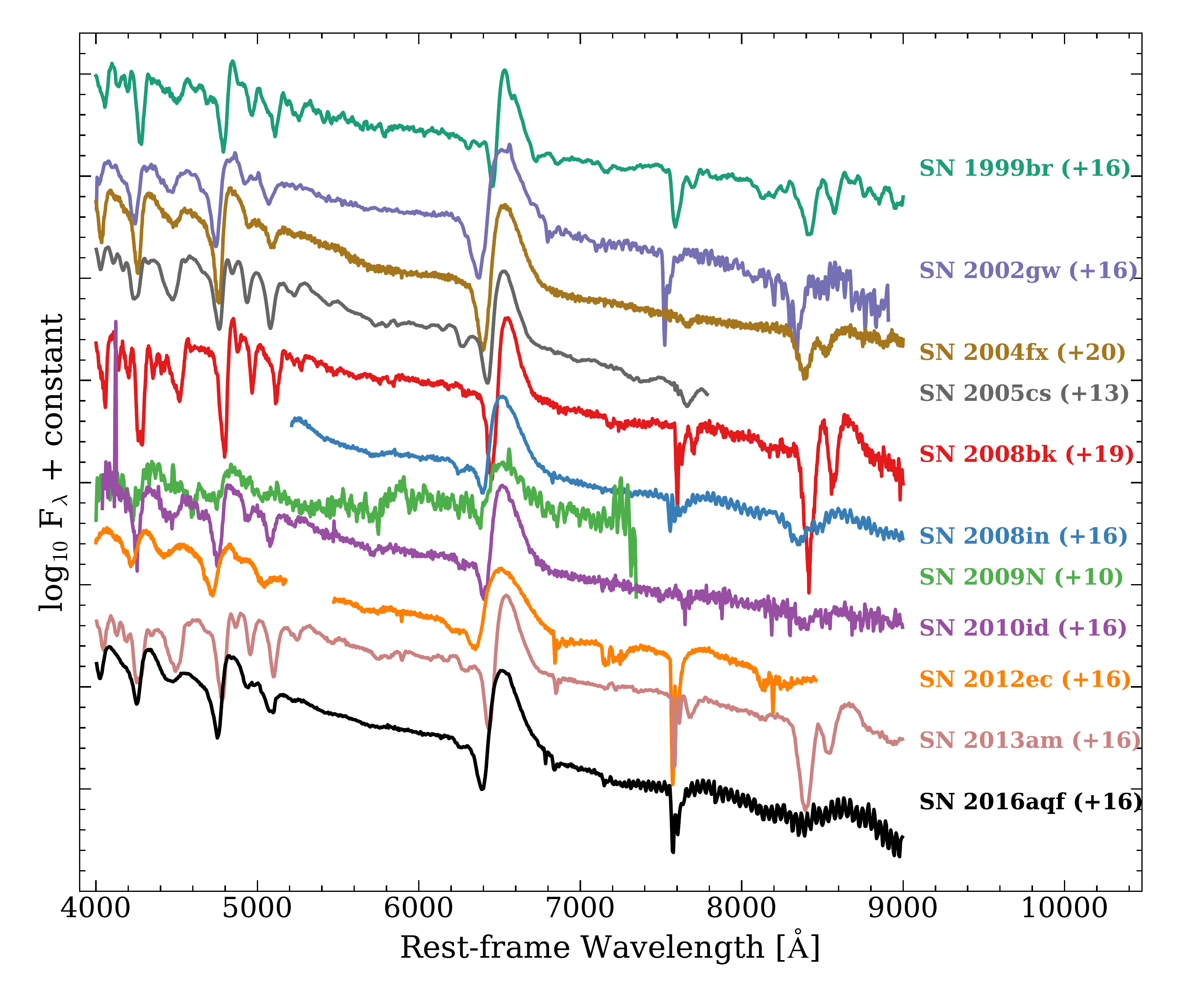}}%\quad
    \subfigure{\includegraphics[scale=0.28]{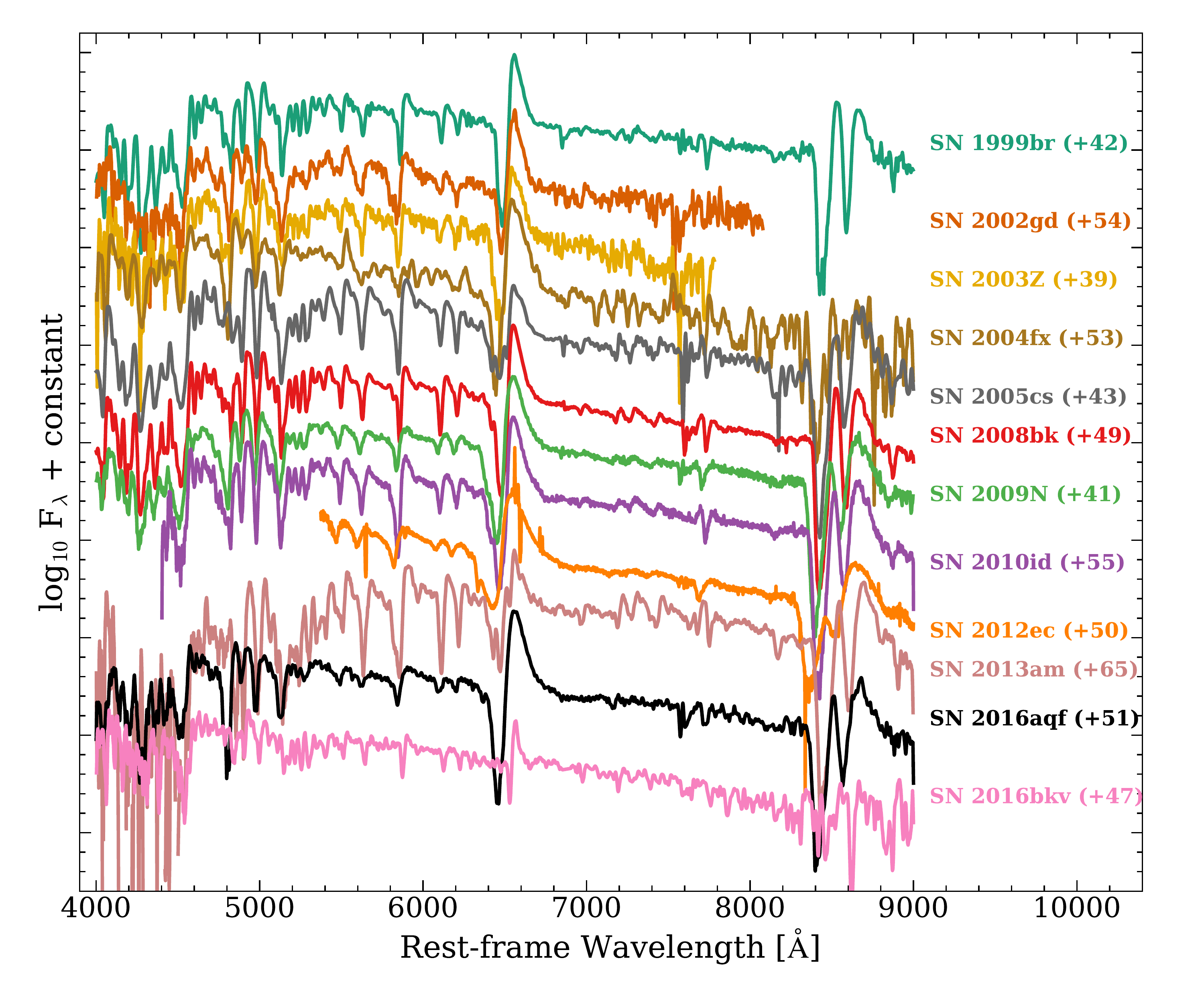}}
    \caption{\sn\ spectrum around +15\,d (\textit{left}) and +50\,d (\textit{right}) compared with the LL~SNe~II sample at similar epochs. Spectra corrected for MW and host galaxy extinction (except for those with values reported as upper limits, see Table \ref{tab:sn_sample}).}
    \label{fig:early_spectra}
\end{figure*}

Fig.~\ref{fig:early_spectra} shows the spectra of \sn\  with other SNe from our comparison sample. The \ion{Fe}{ii} lines are present in all SNe, although in \sn\ they are generally weaker. \sn\ is similar to SN\,2002gw and SN\,2010id, with a relatively featureless spectrum between H$\beta$ and H$\alpha$. However, we see no major differences with the rest of the sample at $\sim$\,+15\,d.

At around +50\,d (Fig.~\ref{fig:early_spectra}), \sn\ resembles SN\,2009N, with the difference that the \ion{Sc}{ii}/\ion{Fe}{ii}~$\lambda5531$, \ion{Sc}{ii}~$\lambda5663$, $\lambda6247$ and \ion{Ba}{ii}~$\lambda6142$ lines are weaker (and weaker than most other SNe in our sample). \ion{O}{i} $\lambda7774$ is seen in the spectrum of most SNe, except  SN\,2002gd and SN\,2016bkv where the signal-to-noise/resolution of the spectra precludes a secure identification. Most SNe have very similar \ion{Fe}{ii} and \ion{Ca}{ii}\,NIR line profiles. \sn\ does not display any other peculiarity with respect to the comparison sample. Note that host galaxy extinction may be substantial for SN\,2013am \citep{Zhang14, Tomasella18}, explaining the drop in flux at the bluer end of this SN.

Table~\ref{tab:spectra_photospheric} shows a list of lines with pseudo-Equivalent Width (pEW, not corrected for instrumental resolution), including the full-width at half maximum (FWHM, not corrected for instrumental resolution) of H$\alpha$, measured from the spectra of \sn\ during the optically-thick phase.

\begin{center}
\input{spectra_photospheric.tex}
\end{center}

%%%%%%%%%%%%%%%%%%%%%%%%%%%%%%%%%%%%%%%%%%%%%%%%%%

\subsection{Nebular spectral evolution}
\label{subsec:nebular_spec}

\begin{figure*}
	\includegraphics[width=\textwidth]{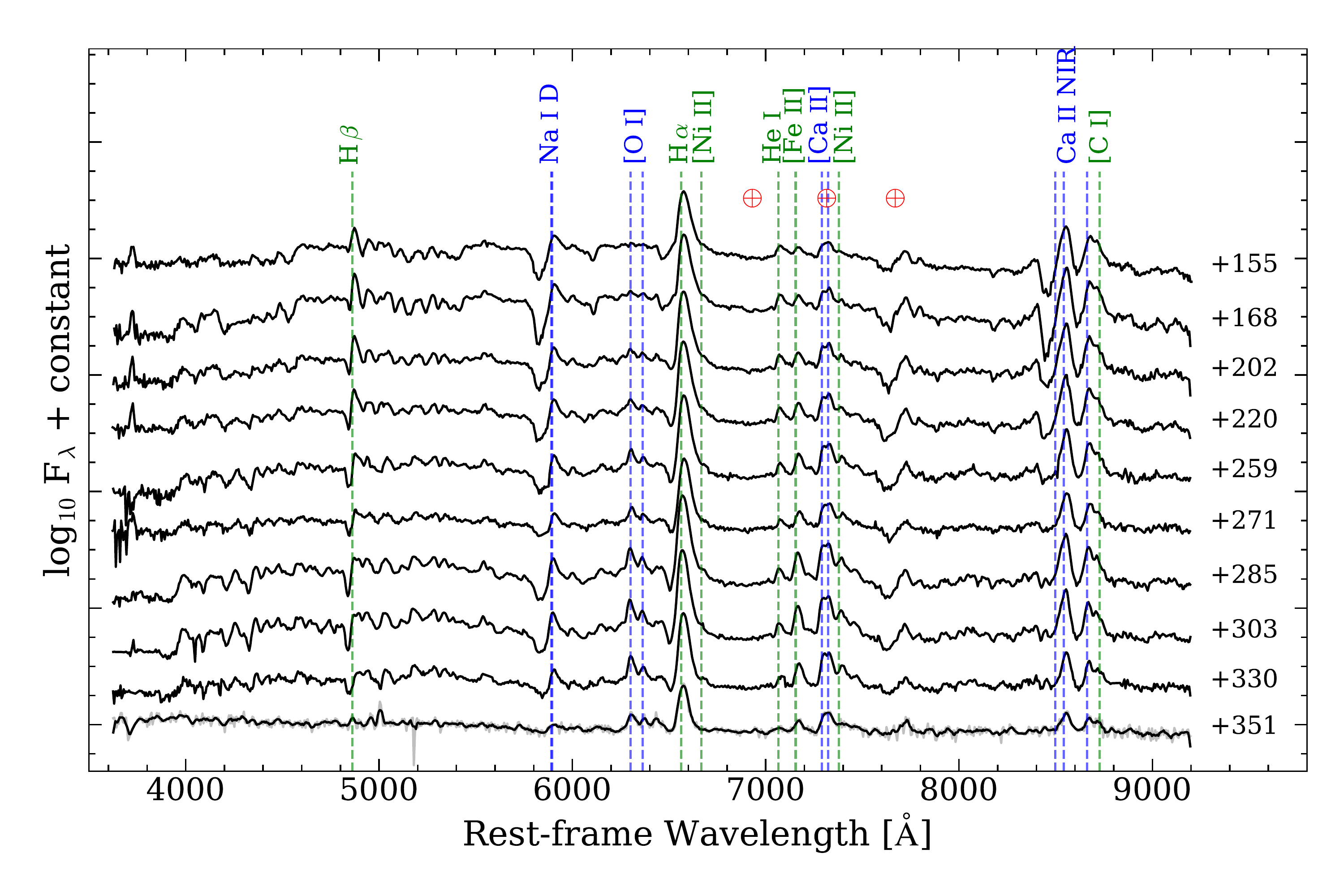}
    \caption{\sn\ nebular phase spectroscopy. H\,$\alpha$, H\,$\beta$, \ion{Na}{i}\,D, [\ion{O}{i}] $\lambda\lambda6300, 6364$, [\ion{Ni}{ii}] $\lambda6667$, \ion{He}{i}~$\lambda7065$, [\ion{Fe}{ii}] $\lambda7155$, [\ion{Ca}{ii}] $\lambda\lambda7291, 7323$, [\ion{Ni}{ii}] $\lambda7378$ and \ion{Ca}{ii}\,NIR lines are shown for guidance. The rest of the description is the same as in Fig.~\ref{fig:photospheric_spectra}.}
    \label{fig:nebular_spectra}
\end{figure*}

Fig.~\ref{fig:nebular_spectra} shows the spectra taken during the optically thin phase. H$\beta$ is present, although its strength slowly decreases at $>250$\,d. The \ion{Fe}{ii} lines around $5000$\,\AA\ are weak and hard to distinguish.
%The \ion{Na}{i}\,D absorption slowly becomes weaker until it almost disappear at +327\,d.
The [\ion{O}{i}] $\lambda\lambda6300, 6364$ doublet has two distinguishable components (separated by $\sim 62$\,\AA), and appears after five months, becoming prominent. At five months, we see the presence of \ion{He}{i}~$\lambda7065$, [\ion{Fe}{ii}] $\lambda7155$, [\ion{Ca}{ii}] $\lambda\lambda7291, 7323$ and [\ion{Ni}{ii}] $\lambda7378$, which become prominent at later epochs. Despite being a LL~SN~II, \sn\ displays blended [\ion{Ca}{ii}] $\lambda\lambda7291, 7323$ lines. The presence of \ion{O}{i} $\lambda7774$ is more prominent at these later epochs. The \ion{Ca}{ii}\,NIR lines are easy to distinguish given the narrow profiles. 

The [\ion{O}{i}] $\lambda\lambda6300, 6364$ and [\ion{Ca}{ii}] $\lambda\lambda7291, 7323$ lines show some very minor redshift ($\sim 5$ \AA, or $\sim 230$ km\,s$^{-1}$ and $\sim 200$ km\,s$^{-1}$), while the \ion{He}{i}~$\lambda7065$, [\ion{Fe}{ii}] $\lambda7155$ and [\ion{Ni}{ii}] $\lambda7378$ lines are more redshifted ($\sim$\,15 \AA, or $\sim 630$ km\,s$^{-1}$, $\sim 630$ km\,s$^{-1}$ and $\sim 610$ km\,s$^{-1}$) throughout most of the nebular phase. We also noticed that the [\ion{Ni}{ii}] $\lambda7378$ line shows almost no redshift ($\sim$\,2 \AA, or $\sim 80$ km\,s$^{-1}$) at $\sim$\,+150 days before rapidly increasing to $\sim$\,10 \AA\ ($\sim 400$ km\,s$^{-1}$) at $\sim$\,+165 days and $\sim$\,20 \AA\ ($\sim 800$ km\,s$^{-1}$) at $\sim$\,+270 days. In addition, the [\ion{O}{i}] $\lambda\lambda6300, 6364$ lines show a minor blueshift ($\sim$\,5 \AA, $\sim 230$ km\,s$^{-1}$) at $\sim$\,+280 days and then gets blueshifted again in about one month. These shifts could be caused by asymmetries caused by clumps in different layers of the expanding envelope. It is worth mentioning that the [\ion{Fe}{ii}]~$\lambda7172$ and [\ion{Ni}{ii}]~$\lambda7412$ lines can contribute to the shifts in the [\ion{Fe}{ii}] $\lambda7155$ and [\ion{Ni}{ii}] $\lambda7378$ lines, respectively. However, due to the resolution of the spectra, we are unable to discern their contribution. Table~\ref{tab:spectra_nebular} contains a list of lines and FWHM measurements of \sn.

\begin{center}
\input{spectra_nebular.tex}
\end{center}

\begin{figure*}
	\includegraphics[scale=0.5]{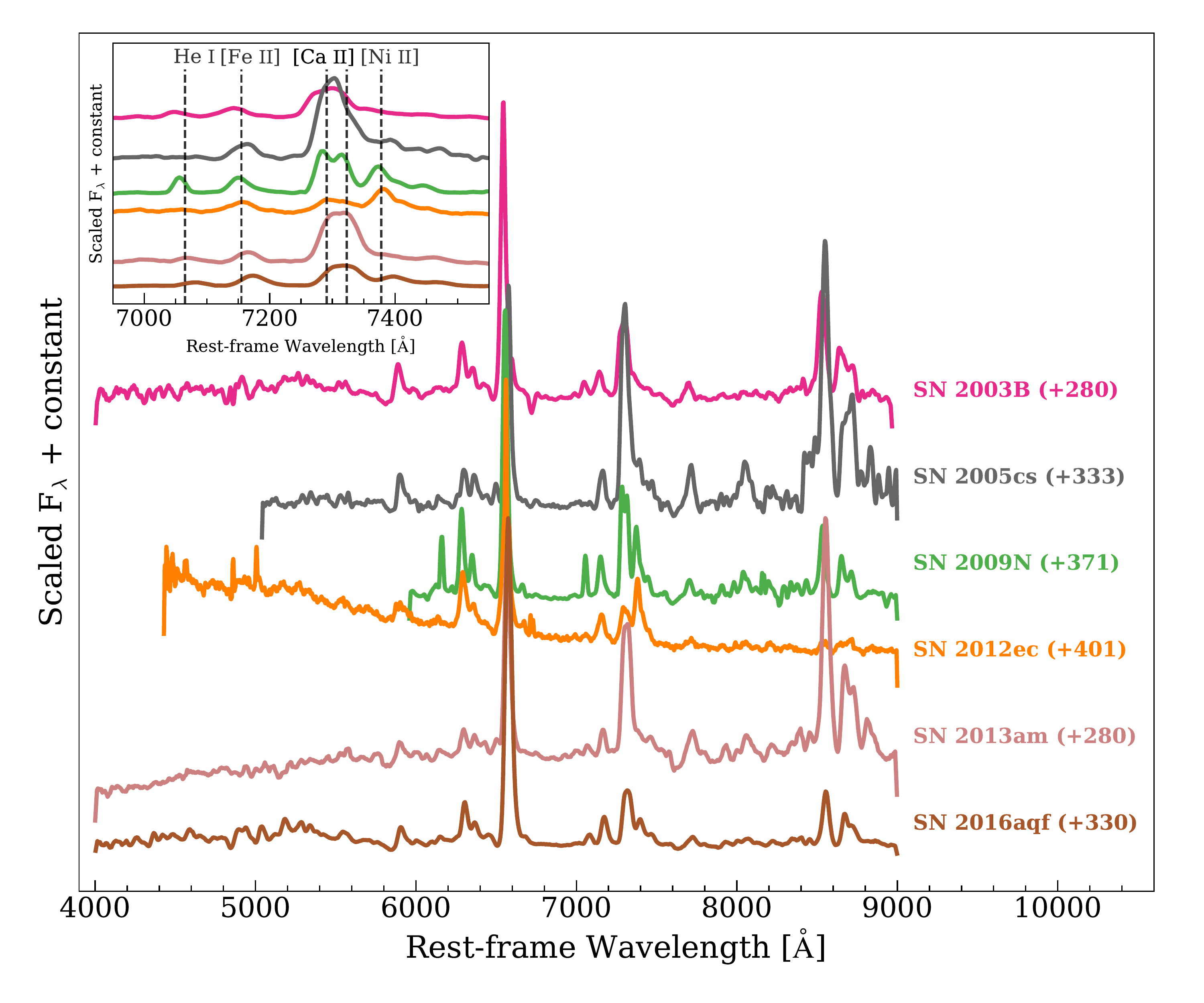}
    \caption{\sn\ spectrum around +330\,d compared with the LL~SN~II sample at similar epochs. The spectra were normalised by their peak H$\alpha$ flux. \textit{Embedded figure}: zoom-in around $\sim 7250$\,\AA. The rest-frame position of the \ion{He}{i}~$\lambda7065$, [\ion{Fe}{ii}]~$\lambda7155$, [\ion{Ca}{ii}] $\lambda\lambda7291, 7323$ and [\ion{Ni}{ii}]~$\lambda7378$ lines are shown. Spectra corrected for MW host galaxy extinction (except for those with values reported as upper limits, see Table \ref{tab:sn_sample}).}
    \label{fig:late_spectra}
\end{figure*}

When we compare \sn\ to other SNe at $>+300$\,d (see Fig.~\ref{fig:late_spectra}), some of them do not show \ion{He}{i}~$\lambda7065$ (e.g., SN\,2005cs and SN\,2012ec). For SN\,2009N, which does show this line, it has a similar strength to [\ion{Fe}{ii}] $\lambda7155$, which does not occur for other SNe. The ratio between the [\ion{O}{i}] $\lambda\lambda$6300, 6364 lines are  similar for all SNe, except for SN\,2005cs where these lines have a similar flux. It can also be seen that [\ion{Ni}{ii}] $\lambda7378$ is easy to distinguish in some SNe (e.g., SN\,2012ec, SN\,2009N and \sn). In the case of SN\,2003B and SN\,2005cs, this line is present, but it gets blended with the [\ion{Ca}{ii}] $\lambda\lambda7291, 7323$ doublet. SN\,2012ec is a special case as it is the only SN that shows a higher peak in [\ion{Ni}{ii}] $\lambda7378$ than in the [\ion{Ca}{ii}] $\lambda\lambda7291, 7323$ doublet.

%%%%%%%%%%%%%%%%%%%%%%%%%%%%%%%%%%%%%%%%%%%%%%%%%%

\subsection{Expansion velocity evolution}
\label{subsec:vel}

The ejecta expansion velocities were measured from the position of the absorption minima for H$\beta$, \ion{Fe}{ii}~$\lambda4924$, \ion{Fe}{ii}~$\lambda5018$, \ion{Fe}{ii}~$\lambda5169$, \ion{Na}{i}\,D (middle of the doublet), \ion{Ba}{ii}~$\lambda6142$, \ion{Sc}{ii}~$\lambda6247$ and H$\alpha$. For H$\alpha$, we also estimated the expansion velocity from the FWHM (corrected for the instrumental resolution) of the emission by using $v = c\times \mathrm{FWHM}/\lambda_{\rm rest}$, where $c$ is the speed of light. We include uncertainties in the measurement of the absorption minima, from the host galaxy recession velocity (3\,km\,s$^{-1}$, as reported in HyperLEDA\footnote{http://leda.univ-lyon1.fr}; \citealt{Makarov14}), the maximum rotation velocity of the galaxy (44.2\,km\,s$^{-1}$, as reported in HyperLEDA) and from the instrumental resolution, all added in quadrature. The major contribution to the uncertainty comes from the instrumental resolution.

\begin{figure}
	\includegraphics[width=\columnwidth]{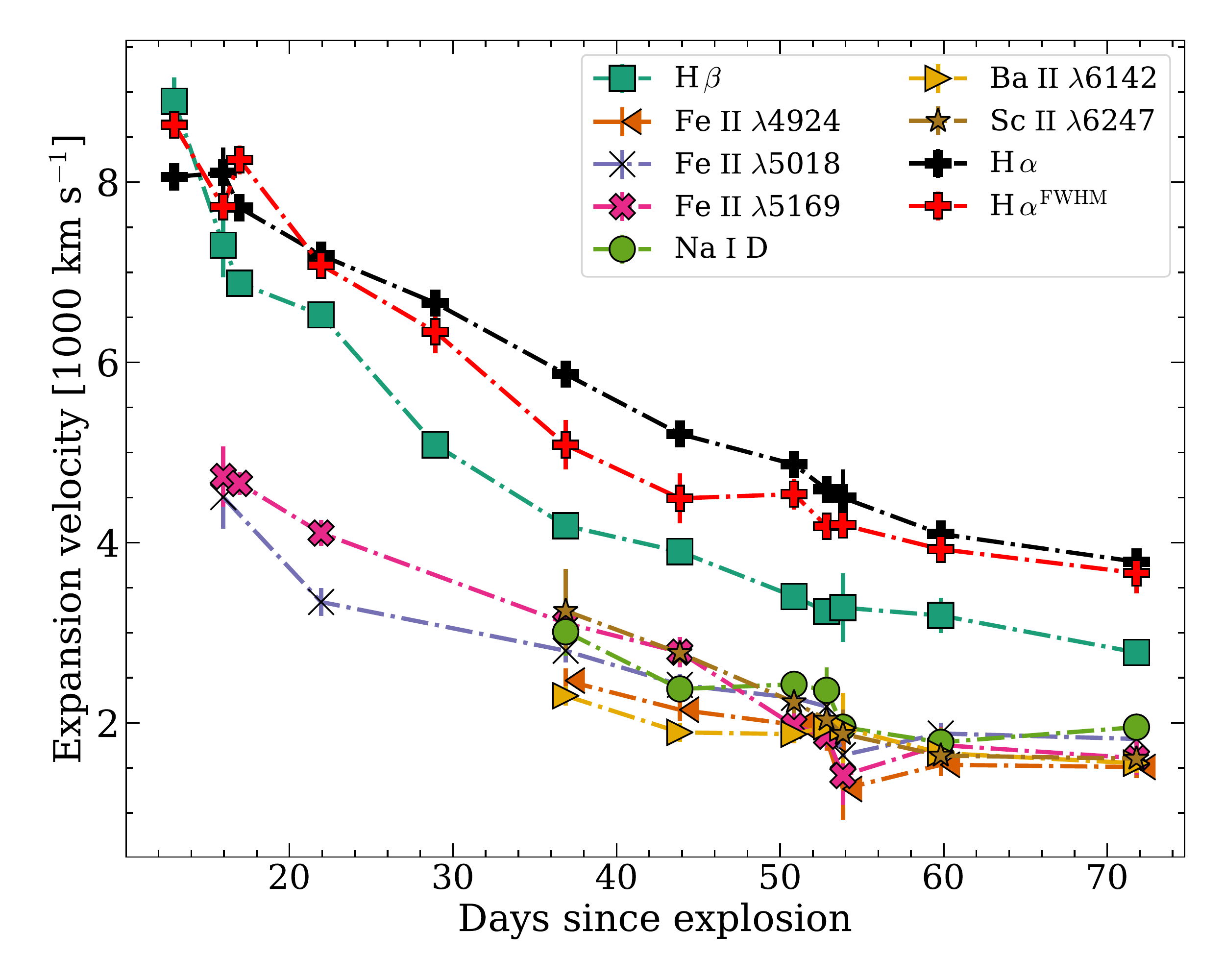}
    \caption{\sn\ expansion velocities for H\,$\beta$, \ion{Fe}{ii}~$\lambda4924$, \ion{Fe}{ii}~$\lambda5018$, \ion{Fe}{ii}~$\lambda5169$, \ion{Na}{i}\,D (middle of the doublet), \ion{Ba}{ii}~$\lambda6142$, \ion{Sc}{ii}~$\lambda6247$ and H\,$\alpha$. For H\,$\alpha$ we also estimated the expansion velocity from the FWHM; see text for details.}
    \label{fig:photosferic_velocities}
\end{figure}

The expansion velocity curves are shown in Fig.~\ref{fig:photosferic_velocities}. The velocities of H$\alpha$ and H$\beta$ are relatively high ($\gtrsim8000$\,km\,s$^{-1}$) at very early epochs ($t\lesssim10$ days) and drop to $\sim5000$ and $4000$\,km\,s$^{-1}$ at $\sim50$ days, respectively, decreasing at a slower rate afterwards. The H$\alpha$ velocity estimated from the FWHM is close to that estimated from the absorption minima as shown by \citet{Gutierrez17a}. The velocities of other lines decrease less dramatically, from $\sim 5000$\,km\,s$^{-1}$ at early epochs ($t\sim$ 10\,d), for the \ion{Fe}{ii} lines, dropping down to $\sim 3000$ km s$^{-1}$ at $\sim$\,50\,d, and then constant thereafter.

In general, the expansion velocity curves of \sn\ fall within the bulk of our sample and follow the general trend, although some of the velocities seem to decrease faster during the first 50 days after explosion.

%%%%%%%%%%%%%%%%%%%%%%%%%%%%%%%%%%%%%%%%%%%%%%%%%%
%%%%%%%%%%%%%%%%%%%%%%%%%%%%%%%%%%%%%%%%%%%%%%%%%%
%%%%%%%%%%%%%%%%%%%%%%%%%%%%%%%%%%%%%%%%%%%%%%%%%%

\section{Physical Parameters}
\label{sec:physical_parameters}

\subsection{Nickel Mass}
\label{subsec:ni_mass}

The \mni\ is one of the main physical parameters that characterise CCSNe as it is formed very close to the core (within a few thousand kilometers; e.g., \citealt[][]{Kasen09}). We estimated the nickel mass of \sn\ by using different methods. These come from: (i) \citet{Arnett96}, (ii) \citet{Hamuy03}, (iii) \citet{Maguire12} and (iv) \citet{Jerkstrand12}. For more information regarding the different relations used for the estimation of the nickel mass, see Appendix \ref{app:nickel_mass_estimation}.

For (i), (ii) and (iv), we used the bolometric luminosity of the exponential decay tail at +200 days, calculated in Sec.~\ref{subsec:lbol} by interpolating with Gaussian Process \citep{Rasmussen06}, using the \textsc{python} package \textsc{george}\footnote{\url{https://github.com/dfm/george}} \citep{Ambikasaran15} and including the distance of the SN for (ii). In the case of (iii), we measured the FWHM of $H_{\alpha}$ at +351 days, correcting it for the FWHM of the instrument. The \mni\ values  obtained with the different methods were M$_{\rm Ni}$ = $0.008^{+0.002}_{-0.002}$, $0.011^{+0.003}_{-0.003}$, $0.014^{+0.009}_{-0.007}$ and $0.007^{+0.001}_{-0.001}$ M$_{\odot}$, respectively. Using the different methods we estimated a weighted mean and a weighted standard error of the mean of M$_{\rm Ni} = 0.008 \pm 0.002$ \msun.

%%%%%%%%%%%%%%%%%%%%%%%%%%%%%%%%%%%%%%%%%%%%%%%%%%

\subsection{Explosion Energy, Ejected Mass and Progenitor Radius}
\label{subsec:physical_parameters}

\citet{Popov93} derived analytical relations for the estimation of the explosion energy (\eexp), ejected envelope mass (\menv) and the progenitor radius prior to outburst (\rprog) for SNe~II-P (following a similar analysis by \citealt{Litvinova85}). These parameters are related to different light-curve properties and also \mni, therefore, they are essential for the characterisation of SNe~II and CCSNe in general. The relations found by \citet{Popov93} are

\begin{equation}
    \log_{10}(\eexp) = 4.0~\log_{10}~t_{\text{p}} + 0.4~M_V + 5.0~\log_{10}(v_{\text{ph}}) - 4.311,
\end{equation}    
\begin{equation}
    \log_{10}(\menv) = 4.0~\log_{10}~t_{\text{p}} + 0.4~M_V + 3.0~\log_{10}(v_{\text{ph}}) - 2.089,
\end{equation}    
and
\begin{equation}
    \log_{10}(\rprog) = -2.0~\log_{10}~t_{\text{p}} - 0.8~M_V - 4.0~\log_{10}(v_{\text{ph}}) - 4.278,
\end{equation}
where $M_V$ is the $V$-band absolute magnitude at the middle of the plateau, $t_{\rm p}$ is the duration of the plateau in days (as in \citealt{Hamuy03}), $v_{\rm ph}$ is the expansion velocity of the photosphere at $t_{\rm p}/2$ (usually measured from the \ion{Fe}{ii}~$\lambda5169$ line, as it has shown to be a good tracer of the photosphere) in $10^3$\,km\,s$^{-1}$. \eexp\ is expressed in $10^{51}$ erg, and \menv\ and \rprog\ in solar units. We measured M$_V = -14.63 \pm 0.27$ mag for which we used Gaussian processes to interpolate the light curve. By using the relativistic Doppler shift, we obtained $v_{\rm ph} = 2068 \pm 167$ km s$^{-1}$ from the \ion{Fe}{ii}~$\lambda5169$ absorption line minima. Finally, we use $t_{\rm p} = 97.9 \pm 7.2$ days, for which we assumed the same value of SN\,2003fb, adding its uncertainty \citep[see][]{Anderson14} in quadrature, as these SNe have relatively similar evolution around the transition ($t\gtrsim$ +50 days) in the $V$ band (see Appendix \ref{app:V-band comparison}). With these values for \sn\ we obtained \eexp\ $= 0.24~\pm~0.13~\times 10^{51}$ erg, \menv\ $= 9.31~\pm~4.26$ \msun\ and \rprog\ $= 152~\pm~94~\text{R}_{\odot}$. The large uncertainties come mainly from the velocity, specifically from the instrumental resolution, and from the distance uncertainty used in calculating the absolute magnitude. We compared these results with similar relations found in the literature \citep[e.g.,][]{Kasen09, Shussman16, Sukhbold16, Kozyreva19, Goldberg19, Kozyreva20}, obtaining similar results.

\sn\ follows the \eexp$-$\mni\ relation found in SNe~II \citep[e.g.,][]{Pejcha15b, Muller17}, and \menv\ follows the \menv$-$\eexp\ relation \citep[e.g.,][]{Pejcha15b}. If we assume a neutron star ($\sim 1.4$\,\msun) as the compact remnant, the progenitor of \sn\ should be a RSG with $\sim 10.7$\,\msun. This is a lower limit, as some mass loss is expected due to various processes, e.g., winds \citep[e.g.][]{Dessart13b}. Finally, \rprog\ is well within the normal values of RSG radii, although on the lower end \citep[e.g.,][]{Pejcha15b, Muller17}, but consistent with other estimations for this sub-class of SN \citep[e.g.,][]{Chugai00, Zampieri03, Pastorello09, Roy11}.

%%%%%%%%%%%%%%%%%%%%%%%%%%%%%%%%%%%%%%%%%%%%%%%%%%
%%%%%%%%%%%%%%%%%%%%%%%%%%%%%%%%%%%%%%%%%%%%%%%%%%
%%%%%%%%%%%%%%%%%%%%%%%%%%%%%%%%%%%%%%%%%%%%%%%%%%

\section{Discussion}
\label{sec:discussion}

%%%%%%%%%%%%%%%%%%%%%%%%%%%%%%%%%%%%%%%%%%%%%%%%%%

\subsection{Progenitor Mass}
\label{subsec:prog_mass}

The progenitors of SNe~II have been extensively studied through pre-SN images \citep[e.g.][]{Smartt09, Smartt15} and hydrodynamical models \citep[e.g.][]{Bersten11, Dessart13b, Martinez19}. Although there remain some disagreements \cite[e.g.,][for discussions of this discrepancy]{Utrobin09, Dessart13b}, there have been recent major improvements due to better cadence observations.

\begin{figure*}
	\includegraphics[width=\textwidth]{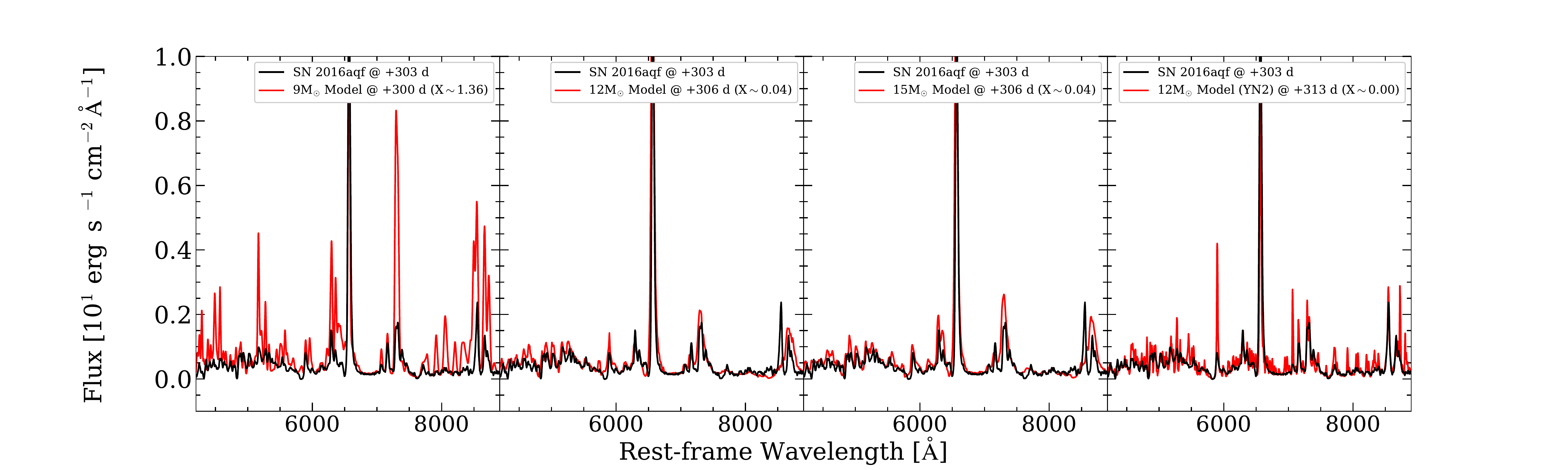}
    \caption{\textit{First three panels (from left to right)}: Spectral synthesis models of SNe~II from \citetalias{Jerkstrand14} and \citetalias{Jerkstrand18}. Three spectral synthesis models at $\sim$+300\,d from different progenitor masses: 9 (left panel), 12 (centre) and 15\,M$_{\sun}$ (right). $X$ is the scaling factor (see Sec.~\ref{subsec:prog_mass}). The 12 and 15\,\msun\ models fit the spectrum better than the 9\,\msun model, including the [\ion{O}{i}] $\lambda\lambda6300, 6364$ lines. \textit{Last panel}: YN2 model of 12\,\msun\ from \citet[][]{Lisakov17}. There is a relatively good agreement with some of the Ca and the [\ion{O}{i}] $\lambda6300, 6364$ lines, however, most other lines are over-predicted.}
    \label{fig:progenitor_models}
\end{figure*}

The [\ion{O}{i}] $\lambda\lambda6300, 6364$ nebular-phase lines have also been shown to be good tracers of the core mass of CCSN progenitors \cite[e.g., ][]{Elmhamdi03, Sahu06, Maguire10}, as at these later epochs we are observing deeper into the progenitor structure. Spectral modelling of the nebular phase has shown good agreement with this and can be used to estimate the progenitor mass \citepalias[e.g., ][]{Jerkstrand12, Jerkstrand14, Jerkstrand18}. In order to estimate the progenitor mass of \sn, we used the spectral synthesis models from \citetalias{Jerkstrand14} and \citetalias{Jerkstrand18} for progenitors with three different ZAMS masses: 9, 12 and 15 M$_{\sun}$. The 9 M$_{\sun}$ model has an initial $^{56}$Ni mass of $0.0062$ \msun\ while the other two models have an initial $^{56}$Ni mass of 0.062 \msun. We compare the nebular spectra of \sn\ with the models at two different epochs each (see Fig.~\ref{fig:progenitor_models} for models at +300 days). The models are scaled by exp((t$_{\rm mod}$ - t$_{\rm SN}$)/111.4), where t$_{\rm mod}$ is the epoch of the spectrum of the models and t$_{\rm SN}$ is the epoch of the spectrum of the SN, by the SN nickel mass, M$_{\rm Ni}^{\rm SN}$/M$_{\rm Ni}^{\rm mod}$, and by the inverse square of the SN distance, (d$_{\rm mod}$/d$_{\rm SN}$)$^2$. The luminosity of some lines, like [\ion{O}{i}] $\lambda6300, 6364$, scale relatively linearly with the \mni\ (as discussed, e.g., in \citetalias{Jerkstrand14}), thus, it is reasonably accurate to compare the models rescaled, with the difference in \mni, to our observed SN. $\chi^2$ values are calculated to quantify these comparisons as well.

From Fig.~\ref{fig:progenitor_models} we see that the 12 and 15 M$_{\odot}$ models present similar results, reproducing several lines. They can partially reproduce the [\ion{O}{i}] $\lambda6300$ line, but the latter does not reproduce the [\ion{O}{i}] $\lambda6364$ line very well. However, these models under-predict the [\ion{Fe}{ii}] $\lambda7155$ line and do not reproduce the [\ion{Ni}{ii}] $\lambda 7378$ line and \ion{Ca}{ii}\,NIR triplet. The 9\,M$_{\sun}$ mostly over-predicts the flux of lines, but does a good job reproducing the \ion{He}{i}~$\lambda7065$ and [\ion{Fe}{ii}] $\lambda7155$ lines. In terms of $\chi^2$ values, the 12 M$_{\sun}$ model is slightly better than the 15 M$_{\sun}$ one, while the 9 M$_{\odot}$ model has a poorer fit. In addition, the 12 M$_{\odot}$ model is relatively consistent with the mass estimate from Sec.~\ref{subsec:physical_parameters}, within the uncertainty. We also measured [\ion{O}{i}]/[\ion{Ca}{ii}] flux ratios \citep[e.g.,][]{Maguire10} between $\sim0.5$--$0.7$, which are consistent with the 12\,\msun\ model and roughly consistent with the 15\,\msun\ model. Finally, we found that the models reproduce lines better at later epochs ($\gtrsim300$\,d) than at early epochs ($<300$\,d). \citetalias{Jerkstrand18} found the same pattern.

There seems to be a very weak detection of [\ion{Ni}{ii}] $\lambda6667$ (see Fig. \ref{fig:nebular_spectra}), partially blended with H\,$\alpha$, and the 9\,\msun\ model predicts similar fluxes for this line and [\ion{Ni}{ii}] $\lambda7378$, due to the high optical depths (fig.~20 of \citetalias{Jerkstrand18}). Note that this model has only primordial nickel in the hydrogen-zone, no synthesised $^{58}$Ni, and a different setup compared to the other two \citepalias[e.g., no mixing applied,][]{Jerkstrand18}. As the model prediction for [\ion{Ni}{ii}] $\lambda7378$ is too weak, one can argue the detection of synthesised nickel. The 9\,\msun\ model over-predicts the [\ion{O}{i}] $\lambda6300, 6364$ lines, including most other lines. As mentioned above, \citetalias{Jerkstrand18} had similar results at these early epochs, however, this model showed better agreement at later epochs (e.g., $>350$\,d for SN\,2005cs). We did not find better agreement at later epochs. 

In order to expand our analysis we also compared \sn\ with the progenitor models from \citet[][]{Lisakov17}, specifically, the YN models of 12\,\msun\ (a set of piston-driven explosion with $^{56}$Ni mixing) as their \mni\ (0.01\,\msun) agree perfectly with our estimation, apart from agreeing with other physical parameters (e.g., \eexp\ = $2.5 \times 10^{50}$\,erg, \menv\ = 9.45\,\msun) as well. This comparison, which was done in the same way as with the other models above, is shown in Fig. \ref{fig:progenitor_models} for the YN2 model as well. As can be seen, the model predicts some of the Ca and the [\ion{O}{i}] $\lambda6300, 6364$ lines relatively well. Nonetheless, most of the other lines are over-predicted. Other models from \citet{Lisakov17} did not show better agreement. However, the fact that both 12\,\msun\ models (from \citetalias{Jerkstrand14} and \citealt{Lisakov17}) partially agree with the [\ion{O}{i}] $\lambda6300, 6364$ lines (the main tracers of the ZAMS mass) strengthen the conclusion that the progenitor is probably a $\sim$ 12\,\msun\ RSG star.

We would like to emphasise that neither the  9\,\msun\ model from \citetalias{Jerkstrand18} nor the YN 12\,\msun\ models from \citet{Lisakov17} have macroscopic mixing. The consistent overproduction of narrow core lines in both models (see Figs. \ref{fig:progenitor_models} and \ref{fig:cmfgen_model}) suggests that mixing is necessary, which the models from \citetalias{Jerkstrand14} have.

%\begin{figure}
%	\includegraphics[width=\columnwidth]{cmfgen_model.pdf}
%    \caption{YN2 model of 12\,\msun\ from \citet[][]{Lisakov17}. There is a relatively good agreement with some of the Ca and the [\ion{O}{i}] $\lambda6300, 6364$ lines, however, most other lines are over-predicted.}
%    \label{fig:cmfgen_model}
%\end{figure}

In conclusion, this shows that the current models have problems predicting the observed diversity of LL SNe II, probably due to the incomplete physics behind these explosions (e.g., assumptions of mixing, $^{56}$Ni mass, rotation). In other words, there is a need of more models with different parameters that can help to understand the observed behaviour of these SNe. As such, we can not exclude a 9\,\msun\ nor a 15\,\msun\ progenitor. Thus, we conclude that the progenitor of \sn\ had a ZAMS mass of 12 $\pm$ 3\,\msun. %, and likely around 12\,\msun. 
A more detailed modelling of the progenitor is needed to improve these constrains, although this is beyond the scope of this work.

%%%%%%%%%%%%%%%%%%%%%%%%%%%%%%%%%%%%%%%%%%%%%%%%

\subsection{He ${\rm I}$ $\lambda7065$}
\label{subsec:heI}

The \ion{He}{i} $\lambda7065$ nebular line has been studied with theoretical modelling (e.g., \citealt{Dessart13a}; \citetalias{Jerkstrand18}), giving a diagnostic of the He shell. These models predict the appearance of this line in SNe~II with low mass progenitors as more massive stars have more extended oxygen shell, shielding the He shell from gamma-ray deposition. However, some LL~SNe~II do not show this line in their spectra (e.g., SN\,2005cs; see Fig.~\ref{fig:late_spectra}). \sn\ shows the clear presence of \ion{He}{i}~$\lambda7065$ throughout the entire nebular coverage. We also see the presence of [\ion{C}{i}] $\lambda8727$, although it gets partially blended with the \ion{Ca}{ii}\,NIR triplet. We expect to see this carbon line as a result of the He shell burning, so the presence of both lines (\ion{He}{i}~$\lambda7065$ and [\ion{C}{i}] $\lambda8727$) is consistent with the theoretical prediction. Thus, we believe that \sn\ is a good case study to provide further understanding of the He shell zone through theoretical models. Furthermore, following the discussion from \citetalias{Jerkstrand18}, we conclude that this is a Fe core SN and not an electron-capture SN (ECSN), as the latter lack lines produced in the He layer.

%%%%%%%%%%%%%%%%%%%%%%%%%%%%%%%%%%%%%%%%%%%%%%%%

\subsection{Ni/Fe abundance ratio}
\label{subsec:nife_ratio}

As discussed above, the nebular spectra of SNe~II contain a lot of information regarding the progenitors as we are looking deeper into its structure. \citetalias[][]{Jerkstrand15a} discussed the importance of the ratio between the [\ion{Ni}{ii}] $\lambda7378$ and [\ion{Fe}{ii}] $\lambda7155$ lines as indicator of the Ni/Fe abundance ratio. These elements are synthesised very close to the progenitor core and, for this reason, their abundances get affected by the inner structure of the progenitor and the explosion dynamics. More specifically, iron-group yields are directly affected mainly by three properties: temperature, density and neutron excess of the fuel \citepalias[for a more detailed account, see][]{Jerkstrand15b}. For this reason, studying iron-group abundances is key to understanding SNe~II.

\begin{figure*}
	\includegraphics[width=\textwidth]{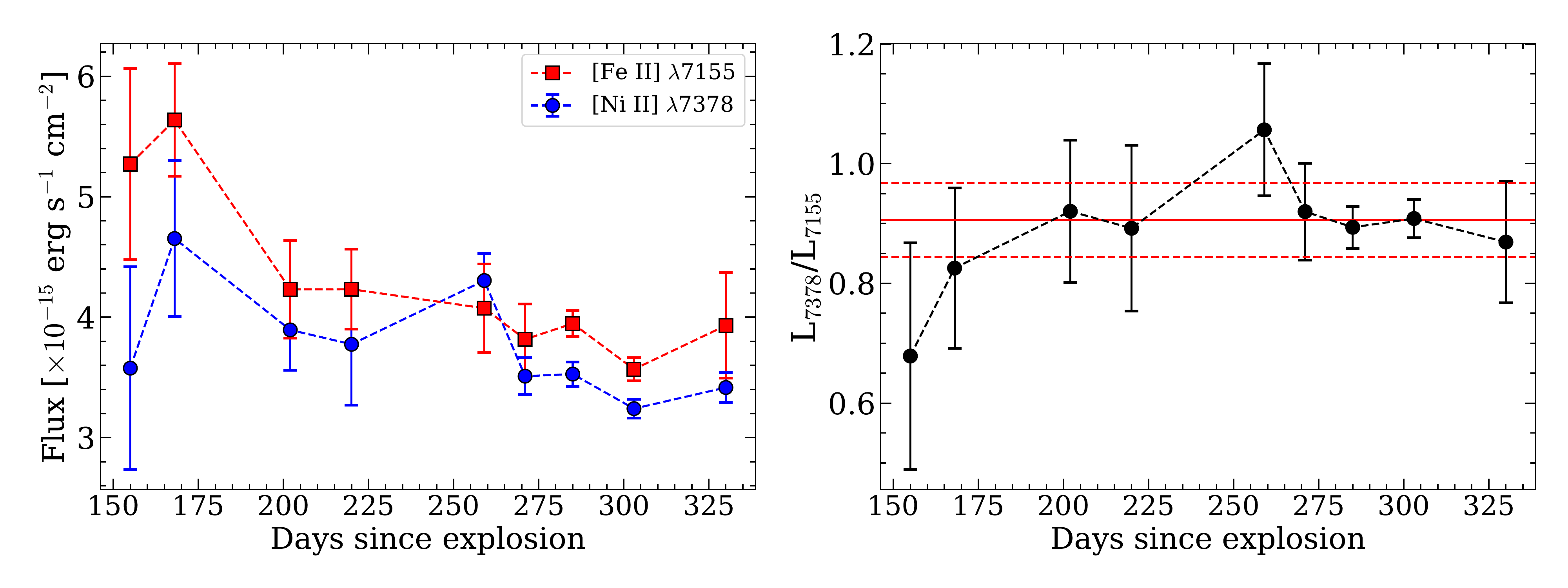}
    \caption{\textbf{Left panel:}[\ion{Ni}{ii}] $\lambda7378$ and [\ion{Fe}{ii}] $\lambda7155$ lines fluxes. \textbf{Right panel:} Luminosity ratio of these lines. The weighted average (solid red line) with a one standard deviation (dashed red lines) are shown for guidance. The value at $\sim$\,+150 days was removed for these calculations.}
    \label{fig:lines_ratio}
\end{figure*}

\sn\ is the only SN~II to date with a relatively extensive coverage of the evolution of [\ion{Ni}{ii}] $\lambda7378$ (most other SNe with the presence of this line only have at most $\sim$\,2 epochs showing it). In Fig.~\ref{fig:lines_ratio} we show the evolution in time of the flux of [\ion{Ni}{ii}] $\lambda7378$ and [\ion{Fe}{ii}] $\lambda7155$, and their luminosity ratio. We estimated the fluxes by fitting Gaussians to the profiles. Uncertainties were estimated by repeating the measurements and assuming different continuum levels, but we do not include the uncertainty coming from the instrumental resolution in any of the measured fluxes throughout this work. However, this should not greatly affect the measurements as the spectral lines are in general much wider than the instrumental resolution (e.g., [\ion{Fe}{ii}] $\lambda7155$ has an average FWHM of $\sim$ 35 \AA).

We notice that the evolution of the luminosity ratio reaches a quasi-constant value after $\sim$\,170 days since the explosion. This suggests that at relatively late nebular phase the Ni/Fe abundance ratio is constant as the temperature should not vary much \citepalias[see][]{Jerkstrand15a}, although clumps in the ejecta might cause deviations from the measured values. After removing the value at $\sim$\,+155 days (as the SN might still be in the transition to the optically thin phase) we report a Ni/Fe luminosity ratio weighted mean of $0.906$ and a standard deviation of $0.062$. The standard deviation gives us a more conservative estimation of the uncertainty in the Ni/Fe luminosity ratio than the uncertainty in the weighted mean.

We follow \citetalias{Jerkstrand15a} to estimate the \ion{Ni}{ii}/\ion{Fe}{ii} ratio and in turn the Ni/Fe abundance ratio. From the ratio between the luminosity of the [\ion{Fe}{ii}] $\lambda7155$ line and \mni, we then obtained a temperature constrain of $T=3919^{+215}_{-257}$\,K. With these values we estimated the Ni/Fe abundance ratio to be $0.081^{+0.009}_{-0.010}$, or $\sim$\,1.4 times the solar ratio ($0.056$, \citealt{Lodders03}). 

However, there are several things we need to take into consideration. Contribution to the [\ion{Fe}{ii}] $\lambda7155$ and [\ion{Ni}{ii}] $\lambda7378$ lines does not come only from synthesised material, but also from primordial Fe and Ni in the H-zone \citepalias{Jerkstrand15a}. The contribution can be significant ($\sim$\,40 per cent) and depends on the model and epoch. Unfortunately, the effect of primordial contamination is not easy to remove without detailed theoretical modelling. Nonetheless, it is plausible that the [\ion{Fe}{ii}] $\lambda7155$ and [\ion{Ni}{ii}] $\lambda7378$ lines are greatly dominated by synthesised Fe and Ni at relatively early epochs ($\lesssim300$\,d), although we are uncertain at which epochs the effect from primordial Fe and Ni starts becoming important \citepalias{Jerkstrand18}. The line ratio can also be affected at very early epochs ($\lesssim200$\,d), as the SN can still be during the optically-thick phase when opacity plays an important role.

Few other SNe have been reported to show [\ion{Ni}{ii}] $\lambda7378$. It is possible that this line is mainly visible in LL~SNe~II, where the expansion velocities are lower, producing narrower deblended line profiles. However, it is also seen in non-LL~SNe~II, other CCSNe \citep[e.g., SN\,2006aj;][]{Maeda07, Mazzali07} and type Ia SNe \citep[SNe~Ia; e.g.][]{Maeda10a}. We searched for objects in our LL~SN~II comparison sample with spectra in which we could detect [\ion{Fe}{ii}] $\lambda7155$ and [\ion{Ni}{ii}] $\lambda7378$ to measure the Ni/Fe abundance ratio as for \sn. We also expanded this sample to include other LL~SNe~II: SN\,1997D, SN\,2003B, SN\,2005cs, SN\,2008bk, SN\,2009N and SN\,2013am.

SN\,1997D and SN\,2008bk were not included in our initial sample as they lack good publicly available data. We also include SN\,2012ec as it is a well-studied case. In the case of SN\,1997D, we measured the ratio at two different epochs, but we used one (at $\sim$\,+384 days) of those, given that the other value (at $\sim$\,+250 days) had relatively large uncertainties. For SN\,2009N we took an average between the two values (at $\sim$\,+372 and +412 days) we were able to measure as they were relatively similar. SN\,2016bkv was not included as the \mni\ values obtained in \citet{Nakaoka18} and \citet{Hosseinzadeh18} for this SN are not consistent with each other ($\sim$\,0.01 \msun\ and 0.0216 \msun, respectively), this being necessary for an accurate estimation of the Ni/Fe abundance ratio. For the rest of the SNe, only one value was obtained. Several other LL SNe II show the presence of [\ion{Ni}{ii}] $\lambda$7378, but it is either blended with other lines or the SNe lack some of the parameters needed to estimate the Ni/Fe abundance ratio.

To expand our analysis we looked into other physical parameters related to the Ni/Fe abundance ratio. For example, \citetalias{Jerkstrand15b} further analyse and compare this ratio against theoretical models. Some of these models show that at lower progenitor mass, the Ni/Fe abundance ratio should be higher. We investigate this by increasing our sample. Unfortunately not many LL~SNe~II have measured progenitor masses from pre-SN images, so we added non-LL~SNe~II as several of these do \citep[e.g.,][]{Smartt15}, while they also show the presence of [\ion{Fe}{ii}] $\lambda7155$ and [\ion{Ni}{ii}] $\lambda7378$ in their spectra. We do not include SNe with estimates of the progenitor mass from other methods as they depend on more assumptions than the pre-SN images method, making these estimates less reliable. The SNe included are: SN\,2007aa \citep{Anderson14,Gutierrez17a}, SN\,2012A \citep{Tomasella13} and SN\,2012aw \citep{Fraser12}. All these SNe are included in Table~\ref{tab:sn_sample}. For SN\,2007aa we calculated the ejected nickel mass to be M$_{\rm Ni}$ = 0.032 $\pm$ 0.009 \msun\ (we estimated this value using the relation from \citealt{Hamuy03} and other values from \citealt{Anderson14}) and estimated the Ni/Fe abundance ratio also as part of this work. For the other two SNe~II, we took the values from \citetalias{Jerkstrand15a}, assuming upper and lower uncertainties equal to the average of the uncertainties of the rest of the sample (not taking into account the uncertainties of SN\,2012ec as they are too high). The Ni/Fe abundance ratio values for this sample are shown in Table~\ref{tab:nife}.

\begin{center}
\input{nife_ratio.tex}
\end{center}

In addition, we compared the Ni/Fe against other physical, light-curve and spectral parameters to investigate possible correlations. The motivation is two-fold. Firstly, we are searching for correlations that might allow indirect methods of measuring this ratio for SNe with blended lines. Secondly, these correlations could shed light on the effect of different parameters in the observed value of Ni/Fe, as is expected for the progenitor mass, important for the theoretical modelling of SNe~II.

\begin{figure*}
	\includegraphics[width=13cm]{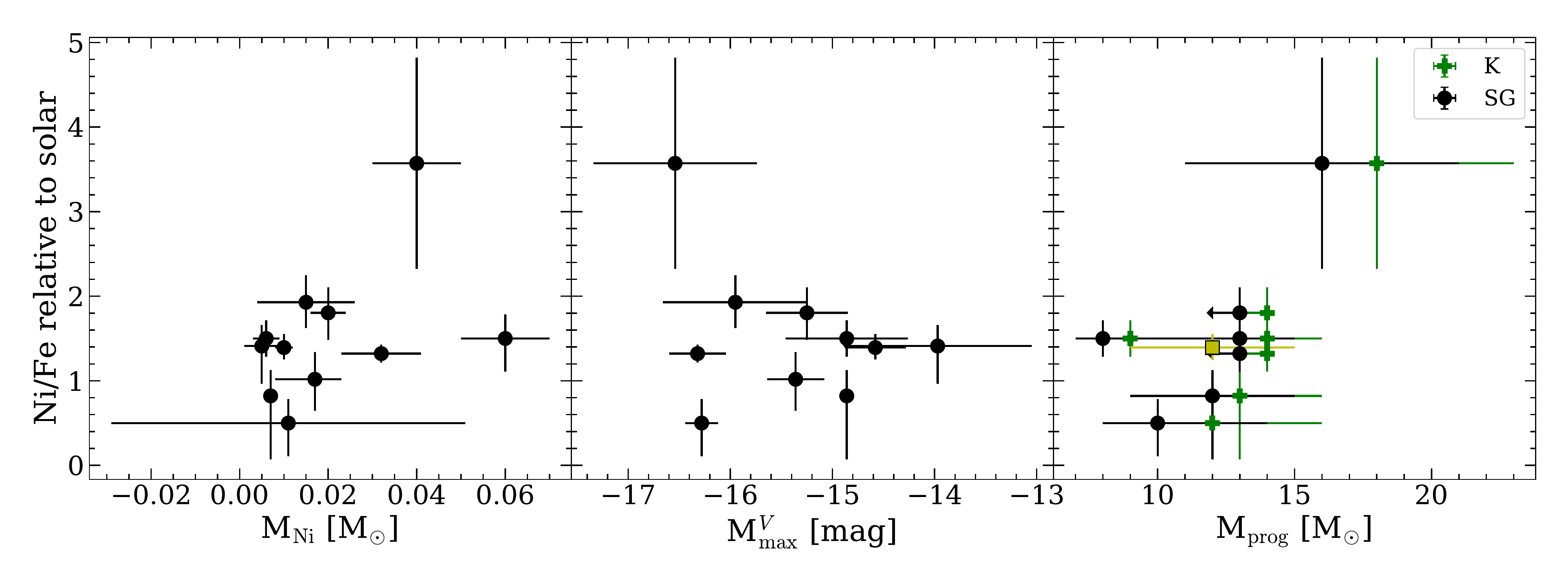}
    \caption{Ni/Fe abundance ratio versus \mni, M$_{\rm max}^V$ and M$_{\rm prog}$. For M$_{\rm prog}$ we show two different progenitor models, KEPLER (K) and STARS and Geneva (SG). \sn\ is shown as a yellow square in the subplot with M$_{\rm prog}$, for which we assume a value of $12 \pm 3$ \msun.}
    \label{fig:line_ratio_analysis}
\end{figure*}

We examined various parameters we thought could be somehow connected to the Ni/Fe abundance ratio. The most relevant parameters are: nickel mass (\mni); SN $V$-band maximum absolute magnitude ($M_{\rm max}^V$); optically-thick phase duration (OPT$_{\rm d}$); \ion{Fe}{ii}~$\lambda5169$ expansion velocity (vel(\ion{Fe}{ii} 5169)); the progenitor mass from KEPLER (K) models (M$_{\rm prog}^{\rm K}$; see \citealt{Smartt15}), progenitor mass from STARS and Geneva (SG) rotating models (M$_{\rm prog}^{\rm SG}$; see \citealt{Smartt15}); explosion energy ($E_{\rm exp}$); [\ion{O}{i}] $\lambda6300$ and [\ion{O}{i}] $\lambda6364$ luminosities at the epoch of measured Ni/Fe abundance ratio ($L_{6300}$ and $L_{6364}$); and host galaxy gas-phase metallicity ((12 + log(O/H))$_{\rm N2}$). The Ni/Fe abundance ratio versus \mni, M$_{\rm max}^V$ and M$_{\rm prog}$ are shown in Fig.~\ref{fig:line_ratio_analysis}.

Pearson and Spearman's rank correlations were used to investigate if there is any meaningful correlation between these parameters and the Ni/Fe abundance ratio. To account for the measurement uncertainties, we use a Monte Carlo method, assuming Gaussian distributions for symmetric uncertainties, skewed Gaussian distributions for asymmetric uncertainties, and a uniform distribution (with a lower limit of 8\,\msun) for upper limits in the progenitor masses.

We found no significant correlation between the parameters tested above. However, we note that the uncertainties in some parameters are significant. If we do not take into account the uncertainties we obtain a weak correlation between Ni/Fe and M$_{\rm max}^V$ and progenitor mass. However, these are mainly driven by one object (SN\,2012ec). 

This null result raises some interesting questions. We did not find a correlation between \mni\ and Ni/Fe abundance ratio, which is expected as one would assume the production of $^{56}$Ni to track the production of $^{58}$Ni and $^{54}$Fe \citepalias[e.g.,][]{Jerkstrand15b}. We expected to see an anti-correlation between progenitor mass and Ni/Fe abundance ratio, as theory predicts that lower-mass stars have relatively thick silicon shells that more easily encompass the mass cut that separates the ejecta from the compact remnant, ejecting part of their silicon layers, which produces higher Ni/Fe abundance ratios. This is supported by the models from \citet{Woosley95} and \citet{Thielemann96}, but not by those of \citet{Limongi03} which use thermal bomb explosions instead of pistons, as the former two do (see \citetalias{Jerkstrand15b}). Having this in mind, our results either indicate that this anti-correlation can be driven by the exact choice of explosion mechanism (e.g., piston-driven explosions, neutrino mechanism, thermal bomb) and physical parameters (e.g., mass cut, composition, density profile), or that low-mass stars typically do not burn and eject Si shells, but either O shells or possibly merged O-Si shells \citep[e.g.,][]{Collins18}. This is an important constraint both for pre-SN modelling (shell mergers and convection physics that determines whether these Si shells are thin or thick) and explosion theory (which matter falls into NS and which is ejected). Finally, we also need to consider the possibility of having primordial Ni and Fe contaminating the measured Ni/Fe abundance ratio, which could affect our results (as discussed above).

%We note that the models shown in fig.~7 of \citetalias{Jerkstrand15b} do not fit the values found here. For example, our data do not support the predicted Ni/Fe dependency on ZAMS masses in the \citet{Woosley95} grid, perhaps suggesting that the mass cuts on the low-mass end in that grid, which were set inside the silicon shells, are too deep, and that instead low-mass progenitors appear to burn and eject oxygen shell material while the silicon shells fall into the neutron star. \red{However, as discussed above, many other parameters can play an important role in this. In addition, as it was already mentioned,} we need to consider the possibility of having primordial Fe and Ni contaminating the synthesised Ni/Fe abundance ratios of some of these SNe, which could affect the analysis.

As mentioned in \citetalias{Jerkstrand15b}, 1D models tend to burn and eject either Si shell or O shell material that gives Ni/Fe abundance ratios of $\sim$\,3 and  $\sim$\,1 times solar, respectively. Therefore, there is a clear-cut prediction that we should see a bimodal distribution of this ratio, with relatively few cases where the burning covers both shells. However, the observed distribution of our sample seems to cover the whole $\sim$\,1--3 range. This may suggest that the 1D picture of progenitors is too simplistic. Recent work on multi-D progenitor simulations (e.g., \citealt{Muller16, Collins18, Yadav20}, and references therein), where some of these suggest vigorous convection and shell mixing inside the progenitor. If this happens, Si and O shells could smear together and burning such a mixture would give rise to Ni/Fe abundance ratios covering the observed range depending on the relative masses of the two components.

%%%%%%%%%%%%%%%%%%%%%%%%%%%%%%%%%%%%%%%%%%%%%%%%%%
%%%%%%%%%%%%%%%%%%%%%%%%%%%%%%%%%%%%%%%%%%%%%%%%%%
%%%%%%%%%%%%%%%%%%%%%%%%%%%%%%%%%%%%%%%%%%%%%%%%%%

\section{Conclusions}
\label{sec:conclusions}

Theoretical modelling has shown that the Ni/Fe abundance ratio, which can be estimated from the [\ion{Ni}{ii}] $\lambda7378$/[\ion{Fe}{ii}] $\lambda7155$ lines ratio, gives an insight of the inner structure of progenitors and explosion mechanism dynamics. To date, very few SNe~II have shown these lines in their spectra, most of them been LL~SNe~II. This could be due to their lower explosion energies (hence lower expansion velocities) which facilitates the deblending of lines, although these lines have also been found in one SN~Ic and SNe~Ia.

\sn\ has a similar spectral evolution to other SNe of this faint sub-class and has a bolometric luminosity and expansion velocities that follow the bulk behaviour of LL~SNe~II. When comparing its nebular spectra to spectral synthesis models to constrain the progenitor mass through the [\ion{O}{i}] $\lambda\lambda 6300, 6364$ lines, we find a relatively good agreement with progenitors of 12 (using two model grids) and 15 \msun. However, due to uncertainties (e.g., mixing) in the other models, we cannot exclude lower mass ($\sim$ 9 \msun) progenitors. In addition, we noted that the lack of macroscopic mixing seen in some models produce too much fine structure in the early nebular spectra, which would need to be considered in future modelling. Hence, we conclude that the progenitor of \sn\ had a ZAMS mass of 12 $\pm$ 3\,\msun. To further constraint the progenitor mass a more detailed modelling would be required, although this is outside the scope of this work.

As observed from the theoretical modelling of SNe II progenitors, the presence of \ion{He}{i}~$\lambda7065$ and [\ion{C}{i}] $\lambda8727$ in the spectra is linked to the (at least partial) burning of the He shell, which would suggest that \sn\ is a Fe-core SN instead of an ECSN.

\sn\ is a unique case as it has an extended spectral coverage showing the evolution of [\ion{Ni}{ii}] $\lambda7378$ and [\ion{Fe}{ii}] $\lambda7155$ lines for over 150 days. The ratio between these lines appears to be relatively constant (at $t\gtrsim$ +170 days), which would suggest that one spectrum at a relatively late epoch would be enough to measure this quantity. An optimal epoch range to measure this ratio is $\sim$\,200--300 days, given that at earlier epochs the SN can still be in the optically-thick phase when the high opacity blocks the contribution from the lines, and at later epochs the contribution from primordial Fe and Ni is more important. This could vary from SN to SN, so a larger sample with extensive coverage of the [\ion{Ni}{ii}] $\lambda7378$ and [\ion{Fe}{ii}] $\lambda7155$ lines is required. When comparing to a sample of SNe~II (LL and non-LL included) with measured Ni/Fe abundance ratio, the \sn\ value falls within the middle of the distribution.

We did not find any anti-correlation between ZAMS mass and Ni/Fe abundance ratio as predicted by theory. We believe this could mean one of two things. On the one hand, as some models predict this anti-correlation, but others do not, this trend could be driven by the choice of explosion mechanism (e.g., piston-driven explosions, neutrino mechanism, thermal bomb) and physical parameters (e.g., mass cut, composition, density profile). On the other hand, this could mean that low-mass stars typically do not burn and eject Si shells, but instead O shells or possibly merged O-Si shells which would alter the produced Ni/Fe abundance ratio. However, one must keep in mind that there is the possibility of having contamination of primordial Ni and Fe, which can be significant (up to $\sim$ 40 per cent) and epoch dependent.

The current picture of 1D progenitors may be too simplistic, as higher dimensional effects, like mixing and convection, can play an important role, which could help reproduce the observed distribution of Ni/Fe abundance ratio.

Finally, we note that nebular-phase spectral coverage of SNe~II is essential for the study of these objects. While there exist a number of SN~II nebular spectra in the literature, additional higher cadence and higher signal-to-noise observations are required to help improve theoretical models.

\section*{Acknowledgements}

This work is based (in part) on observations collected at the European Organisation for Astronomical Research in the Southern Hemisphere under ESO programme 0102.D-0919, and as part of PESSTO, (the Public ESO Spectroscopic Survey for Transient Objects Survey) ESO program 191.D-0935, 197.D-1075. This work makes use of data from Las Cumbres Observatory, the Supernova Key Project, and the Global Supernova Project.

We thank the ASAS-SN collaboration for sharing data of non-detections for this work.

TMB was funded by the CONICYT PFCHA / DOCTORADOBECAS CHILE/2017-72180113.
CPG and MS acknowledge support from EU/FP7-ERC grant No. [615929]. SGG acknowledges support by FCT under Project CRISP PTDC/FIS-AST-31546 and UIDB/00099/2020. L.G. was funded by the European Union's Horizon 2020 research and innovation programme under the Marie Sk\l{}odowska-Curie grant agreement No. 839090. This work has been partially supported by the Spanish grant PGC2018-095317-B-C21 within the European Funds for Regional Development (FEDER).
MG is supported by the Polish NCN MAESTRO grant 2014/14/A/ST9/00121.
MN is supported by a Royal Astronomical Society Research Fellowship.
DAH, GH, and CM were supported by NSF Grant AST-1313484.

This research has made use of the NASA/IPAC Extragalactic Database (NED) which is operated by the Jet Propulsion Laboratory, California Institute of Technology, under contract with the National Aeronautics. We acknowledge the usage of the HyperLeda database (http://leda.univ-lyon1.fr)


\bibliography{References}
\bibliographystyle{mnras}

%%%%%%%%%%%%%%%%%%%%%%%%%%%%%%%%%%%%%%%%%%%%%%%%%%
%%%%%%%%%%%%%%%%% APPENDICES %%%%%%%%%%%%%%%%%%%%%
%%%%%%%%%%%%%%%%%%%%%%%%%%%%%%%%%%%%%%%%%%%%%%%%%%

\appendix

%%%%%%%%%%%%%%%%%%%%%%%%%%%%%%%%%%%%%%%%%%%%%%%%%%%%%%%%%%%%%%%%%%%%%%
%\section{Photometry Table}
%\label{app:phot}

%\begin{center}
%\input{photometry.tex}
%\end{center}

%%%%%%%%%%%%%%%%%%%%%%%%%%%%%%%%%%%%%%%%%%%%%%%%%%%%%%%%%%%%%%%%%%%%%%
%\section{Spectra Information Tables}
%\label{app:spec_tables}

%\begin{center}
%\input{spectra_info.tex}
%\end{center}

%\begin{center}
%\input{spectra_photospheric.tex}
%\end{center}

%\begin{center}
%\input{spectra_nebular.tex}
%\end{center}

%%%%%%%%%%%%%%%%%%%%%%%%%%%%%%%%%%%%%%%%%%%%%%%%%%%%%%%%%%%%%%%%%%%%%%
%\section{SNe~II sample Table}
%\label{app:sn_sample}
%\begin{center}
%\input{sn_sample.tex}
%\end{center}

%%%%%%%%%%%%%%%%%%%%%%%%%%%%%%%%%%%%%%%%%%%%%%%%%%%%%%%%%%%%%%%%%%%%%% 
%\iffalse
\section{Nickel Mass estimation}
\label{app:nickel_mass_estimation}

%%%%%%% Arnet96
In the literature there are various methods to estimate the \nickel mass. These are as follows.

\citet{Arnett96} gives the following relation using SN\,1987A as comparison:
\begin{equation}
    M_{\rm Ni} = 0.075 \times \frac{L_{\text{SN}}}{L_{\text{87A}}}\text{M}_{\odot}
\end{equation}
By using the bolometric light curve calculated in Section \ref{subsec:lbol}, interpolating with Gaussian Processes to obtain the luminosity at 200 days after the explosion, we obtain $M_{\rm Ni} = 0.008 \pm 0.001$ \msun.

%%%%%%% Hamuy03
\citet{Hamuy03} formed a relation between the bolometric luminosity of the exponential decay tail and the \nickel\ mass. The bolometric luminosity is then given by:
\begin{equation}
\begin{split}
    log_{10}~&L_{\rm tail} = \\
             &\frac{-[m_{V, \text{tail}} - A_{\text{MW}}(V) - A_{\text{Host}}(V) + BC] + 5~log_{10}(D) - 8.14}{2.5}
\end{split}
\end{equation}
where $L_\mathrm{tail}$ is the tail luminosity in erg\,s$^{-1}$ at 200 days after the explosion, $D$ is the distance in cm, $BC$ is a bolometric correction that permits one to transform V magnitudes into bolometric magnitudes, and the additive constant provides the conversion from Vega magnitudes into cgs units. From SN\,1987A and SN\,1999em \citet{Hamuy01T} found that $BC = 0.26 \pm 0.06$. Using the relation found by \citet{Hamuy03} the nickel mass is obtained as follows:
\begin{equation}
    M_{\rm Ni} = (7.866 \times  10^{-44})~L_{\text{tail}}~exp\Big[\frac{(t_{\text{tail}} - t_0)/(1 + z) - 6.1}{111.26}\Big]\text{M}_{\odot}
\end{equation}
from which we obtain $M_{\rm Ni} = 0.011 \pm 0.003$ \msun. 

%%%%%%% Maguire12
\citet{Maguire12} found a relation between the nickel mass and the \halpha\ FWHM  given by
\begin{equation}
    M_{\rm Ni} = A \times 10^{B \times FWHM_{\text{corr}}}\text{M}_{\odot}
\end{equation}
where $B = 0.0233 \pm 0.0041$, $A=1.81^{+1.05}_{-0.68} \times 10^{-3}$ and FWHM$_{\rm corr}$ is the FWHM of H$_{\alpha}$, corrected by the spectral resolution of the instrument, during the nebular phase ($\sim 350 - 550$ days). From this relation, using the FWHM of H\,$\alpha$ from the spectrum at +348 days, we obtain $M_{\rm Ni} = 0.014^{+0.009}_{-0.007}$ \msun, where we used FWHM$_{\rm inst}$ = 21.2 \AA, taken from grism \#13 in EFOSC2 (as given in the ESO website).

%%%%%%% Jerkstrand12
\citetalias{Jerkstrand12} also gives a relation to estimate the nickel mass from the early exponential-decay tail, assuming full trapping, that the deposited energy is instantaneously re-emitted and that no other energy source has any influence, i.e.,
\begin{equation}
L_{^{56}\text{Co}}(t) = 9.92 \times 10^{41} \frac{M_{\text{Ni}}}{0.07M_{\odot}}\Big( e^{-t/111.4~d} - e^{-t/8.8~d} \Big)~\text{erg}~\text{s}^{-1}
\end{equation}
from which we obtain $M_{\rm Ni} = 0.007 \pm 0.001$ \msun.
%\fi
%%%%%%%%%%%%%%%%%%%%%%%%%%%%%%%%%%%%%%%%%%%%%%%%%%%%%%%%%%%%%%%%%%%%%%
\section{\textit{V}-band comparison}
\label{app:V-band comparison}

\begin{figure}
	\includegraphics[width=\columnwidth]{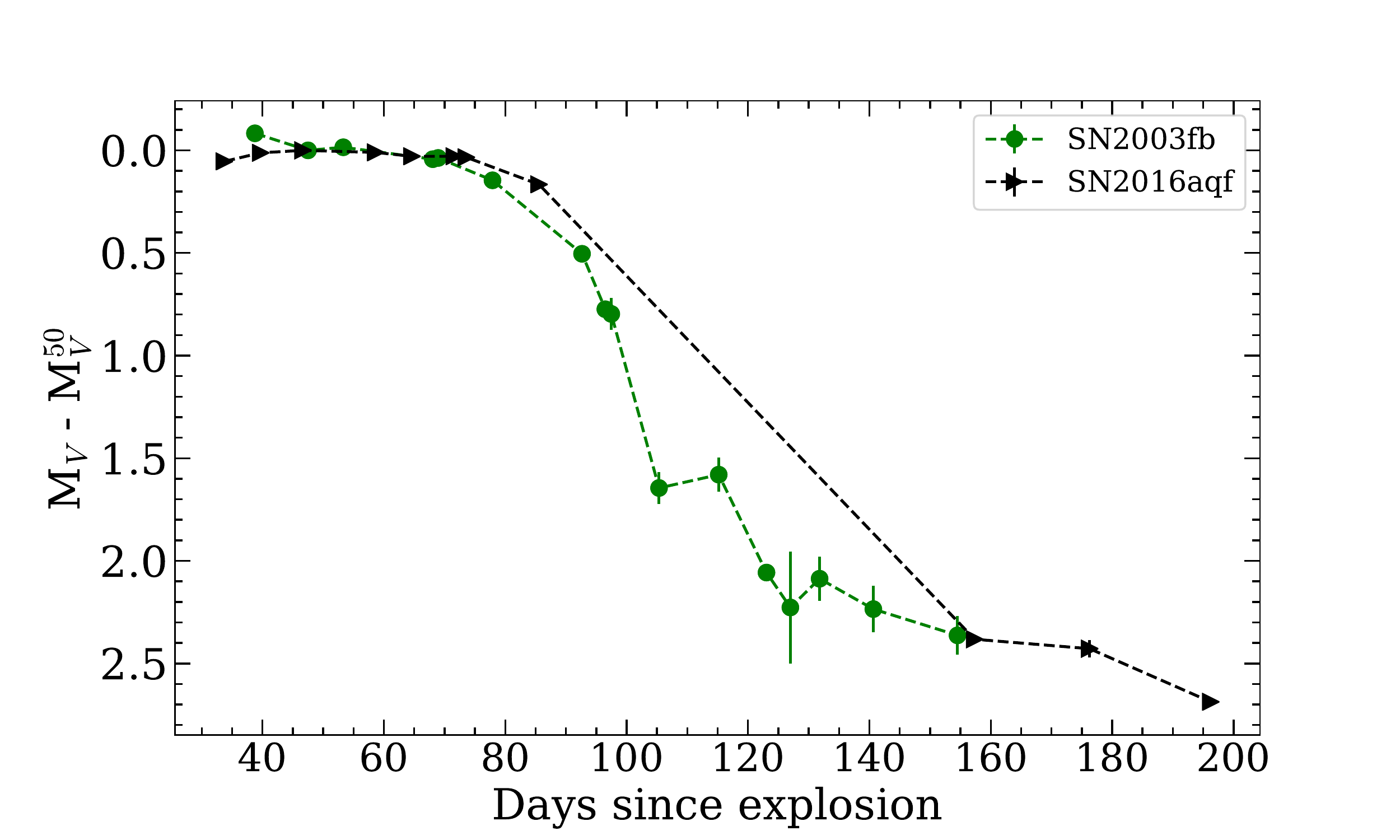}
    \caption{$V$-band comparison between the light curves of \sn\ and SN\,2013fb. The $y$-axis is $V$-band absolute magnitude minus $V$-band absolute magnitude at +50 days. The evolution of both light curves is very similar.}
    \label{fig:V-band_comparison}
\end{figure}

Given that the SN was not visible for a period of time, we do not have observations of the transition from the plateau phase to the nebular phase. To estimate the duration of the plateau, we therefore compared the $V$-band light curve of \sn\ with other LL~SNe~II in our sample. We found that the $V$ band of SN\,2003fb has a similar shape (see Fig.~\ref{fig:V-band_comparison}), if normalised by the luminosity at 50 days after the explosion. For this reason we decided to use the plateau duration of SN\,2003fb (adding its uncertainty in quadrature) for \sn.

\bsp	% typesetting comment
\label{lastpage}
\end{document}

%% file: photometry.tex
\begin{table*}
\centering
\caption{\sn{} $BVgri$-band photometry between +5 and +311 days. $BV$ bands are in Vega magnitude system, while $gri$ bands are in AB magnitude system.}
\begin{tabular}{cccccccccccc}
\hline
MJD   & Phase  & $B$    & $\sigma_B$ & $V$    & $\sigma_V$ & $g$    & $\sigma_g$ & $r$    & $\sigma_r$ & $i$    & $\sigma_i$ \\ \hline
\hline
57448 & 8 & 15.851 & 0.006      & 15.851 & 0.006      & 15.851 & 0.006      & 15.851 & 0.006      & 15.851 & 0.006      \\
57452 & 12 & 15.985 & 0.005      & 15.985 & 0.005      & 15.985 & 0.005      & 15.985 & 0.005      & 15.985 & 0.005      \\
57455 & 15 & 16.057 & 0.010      & 16.057 & 0.010      & 16.057 & 0.010      & 16.057 & 0.010      & 16.057 & 0.010      \\
57456 & 16 & 16.045 & 0.034      & 16.045 & 0.034      & 16.045 & 0.034      & 16.045 & 0.034      & 16.045 & 0.034      \\
57457 & 17 & 16.033 & 0.103      & 16.033 & 0.103      & -      & -          & -      & -          & -      & -          \\
57458 & 18 & 16.095 & 0.039      & 16.095 & 0.039      & 16.095 & 0.039      & 16.095 & 0.039      & -      & -          \\
57459 & 19 & 16.131 & 0.008      & 16.131 & 0.008      & 16.131 & 0.008      & 16.131 & 0.008      & 16.131 & 0.008      \\
57460 & 20 & 16.190 & 0.005      & 16.190 & 0.005      & 16.190 & 0.005      & 16.190 & 0.005      & 16.190 & 0.005      \\
57462 & 22 & 16.268 & 0.031      & 16.268 & 0.031      & 16.268 & 0.031      & 16.268 & 0.031      & 16.268 & 0.031      \\
57468 & 28 & 16.459 & 0.009      & 16.459 & 0.009      & 16.459 & 0.009      & 16.459 & 0.009      & 16.459 & 0.009      \\
57474 & 34 & 16.534 & 0.010      & 16.534 & 0.010      & 16.534 & 0.010      & 16.534 & 0.010      & 16.534 & 0.010      \\
57480 & 40 & 16.576 & 0.008      & 16.576 & 0.008      & 16.576 & 0.008      & 16.576 & 0.008      & 16.576 & 0.008      \\
57487 & 47 & 16.650 & 0.009      & 16.650 & 0.009      & 16.650 & 0.009      & 16.650 & 0.009      & 16.650 & 0.009      \\
57493 & 53 & 16.754 & 0.015      & 16.754 & 0.015      & 16.754 & 0.015      & 16.754 & 0.015      & 16.754 & 0.015      \\
57499 & 59 & 16.812 & 0.014      & 16.812 & 0.014      & 16.812 & 0.014      & 16.812 & 0.014      & 16.812 & 0.014      \\
57505 & 65 & 16.824 & 0.015      & 16.824 & 0.015      & 16.824 & 0.015      & 16.824 & 0.015      & 16.824 & 0.015      \\
57512 & 72 & 16.890 & 0.017      & 16.890 & 0.017      & -      & -          & 16.890 & 0.017      & 16.890 & 0.017      \\
57514 & 74 & 16.912 & 0.012      & 16.912 & 0.012      & 16.912 & 0.012      & 16.912 & 0.012      & 16.912 & 0.012      \\
57523 & 83 & -      & -          & -      & -          & 16.258 & 0.007      & 16.258 & 0.007      & -      & -          \\
57524 & 84 & -      & -          & -      & -          & 16.289 & 0.008      & 16.289 & 0.008      & -      & -          \\
57526 & 86 & 17.182 & 0.030      & 17.182 & 0.030      & 17.182 & 0.030      & 17.182 & 0.030      & 17.182 & 0.030      \\
57598 & 158 & 19.108 & 0.034      & 19.108 & 0.034      & 19.108 & 0.034      & 19.108 & 0.034      & 19.108 & 0.034      \\
57617 & 177 & 19.216 & 0.095      & 19.216 & 0.095      & 19.216 & 0.095      & 19.216 & 0.095      & 19.216 & 0.095      \\

%57637 & 194	& - &	0.039 & 	- & 	0.039 &	- &	0.039 &	- &	0.039 &	- &	0.039\\
%57655 & 212	& - &	0.027 & 	- & 	0.027 &	- &	0.027 &	- &	0.027 &	- &	0.027 \\
%57674 & 231	& - &	0.040 & 	- & 	0.040 &	- &	0.040 &	- &	0.040 &	- &	0.040 \\

57684 & 244 & 20.311 & 0.041      & 20.311 & 0.041      & 20.311 & 0.041      & 20.311 & 0.041      & 20.311 & 0.041      \\
57704 & 264 & 20.430 & 0.074      & 20.430 & 0.074      & 20.430 & 0.074      & 20.430 & 0.074      & 20.430 & 0.074      \\
57726 & 286 & -      & -          & 19.546 & 0.077      & 19.546 & 0.077      & -      & -          & -      & -          \\
57727 & 287 & -      & -          & -      & -          & 19.795 & 0.049      & 19.795 & 0.049      & 19.795 & 0.049      \\
57728 & 288 & 20.709 & 0.063      & 20.709 & 0.063      & 20.709 & 0.063      & 20.709 & 0.063      & 20.709 & 0.063      \\
57749 & 309 & 20.925 & 0.120      & 20.925 & 0.120      & 20.925 & 0.120      & 20.925 & 0.120      & 20.925 & 0.120      \\
57750 & 310 & 20.717 & 0.073      & 20.717 & 0.073      & 20.717 & 0.073      & 20.717 & 0.073      & 20.717 & 0.073      \\
57751 & 311 & 20.932 & 0.076      & 20.932 & 0.076      & 20.932 & 0.076      & 20.932 & 0.076      & 20.932 & 0.076      \\ \hline
\end{tabular}
\label{tab:photometry}
\end{table*}

%% file: spectra_info.tex
\begin{table*}
\centering
\caption{The UTC dates mark the beginning of the exposures. Phase with respect to the explosion epoch (MJD 57440.19).}
\begin{tabular}{lccccc}
\hline
UTC Date                & MJD     & Phase & Range [\AA]  & Resolution [\AA]  & Telescope/Instrument \\ \hline \hline
2016-02-27T09:54:44.475 & 57445.4 & 5     & 3250 - 9300  & 18.0 & FTS/FLOYDS-S         \\
2016-02-27T21:00:25.837 & 57445.9 & 6     & 3600 - 9200  &  7.0 & SALT/RSS             \\
2016-03-01T11:24:32.205 & 57448.5 & 8     & 3250 - 10000 & 18.0 & FTS/FLOYDS-S           \\
2016-03-06T10:09:18.022 & 57453.4 & 13    & 3300 - 10001 & 18.0 & FTS/FLOYDS-S           \\
2016-03-09T04:39:49.731 & 57456.2 & 16    & 3640 - 9235  & 21.2 & NTT/EFOSC2           \\
2016-03-10T12:46:20.372 & 57457.5 & 17    & 3299 - 10000 & 18.0 & FTS/FLOYDS-S           \\
2016-03-15T10:13:25.851 & 57462.4 & 22    & 3250 - 10000 & 18.0 & FTS/FLOYDS-S           \\
2016-03-22T11:10:18.554 & 57469.5 & 29    & 3900 - 9999  & 18.0 & FTS/FLOYDS-S           \\
2016-03-30T09:29:37.393 & 57477.4 & 37    & 3401 - 10000 & 18.0 & FTS/FLOYDS-S           \\
2016-04-06T08:58:10.460 & 57484.4 & 44    & 3299 - 9999  & 18.0 & FTS/FLOYDS-S           \\
%2016-04-06T01:54:06.293 & 57484.1 & 44    & 3645 - 9239  & 21.2 & NTT/EFOSC2           \\
2016-04-13T10:06:55.453 & 57491.4 & 51    & 3599 - 10000 & 18.0 & FTS/FLOYDS-S           \\
2016-04-15T08:39:52.609 & 57493.4 & 53    & 3600 - 10000 & 18.0 & FTS/FLOYDS-S           \\
2016-04-16T00:26:16.687 & 57494.1 & 54    & 3645 - 9239  & 21.2 & NTT/EFOSC2           \\
2016-04-22T08:55:30.146 & 57500.4 & 60    & 3950 - 10000 & 18.0 & FTS/FLOYDS-S           \\
2016-05-04T09:27:45.944 & 57512.4 & 72    & 3650 - 10000 & 18.0 & FTS/FLOYDS-S           \\
2016-07-26T09:49:26.448 & 57595.4 & 155   & 3645 - 9239  & 21.2 & NTT/EFOSC2           \\
2016-08-08T09:39:55.699 & 57608.4 & 168   & 3639 - 9233  & 21.2 & NTT/EFOSC2           \\
2016-09-11T08:20:03.866 & 57642.3 & 202   & 3640 - 9233  & 21.2 & NTT/EFOSC2           \\
2016-09-29T07:23:09.743 & 57660.3 & 220   & 3636 - 9232  & 21.2 & NTT/EFOSC2           \\
2016-11-07T07:57:09.799 & 57699.3 & 259   & 3639 - 9232  & 21.2 & NTT/EFOSC2           \\
2016-11-19T04:31:19.658 & 57711.2 & 271   & 3636 - 9231  & 21.2 & NTT/EFOSC2           \\
2016-12-03T06:56:26.427 & 57725.3 & 285   & 3639 - 9232  & 21.2 & NTT/EFOSC2           \\
2016-12-21T05:55:04.628 & 57743.2 & 303   & 3640 - 9233  & 21.2 & NTT/EFOSC2           \\
2017-01-17T02:55:39.708 & 57770.1 & 330   & 3639 - 9233  & 21.2 & NTT/EFOSC2           \\
2017-02-07T02:46:40.051 & 57791.1 & 351   & 3640 - 9233  & 21.2 & NTT/EFOSC2           \\ \hline
\end{tabular}
\label{tab:spectra_info}
\end{table*}

%% file: sn_sample.tex
\begin{table*}
\centering
\caption{SN II sample used throughout this work. The data for this sample were taken from the references cited in column References.}
\tiny
\setlength\tabcolsep{3.0pt} % default value: 6pt
\begin{tabular}{lcccccccccc}
\hline
SN          & z        & M$_{\rm Ni}$& $\sigma_{\rm Ni}^-$ & $\sigma_{\rm Ni}^+$ & $\mu$& $\sigma_{\mu}$ & $A_V$(MW) & $A_V$(Host) & Host  & References \\
  &        & [M$_{\odot}$] & [M$_{\odot}$]  & [M$_{\odot}$]  & [mag] & [mag] & [mag] &  [mag] &        \\ \hline\hline
SN1997D     & 0.004059 & 0.005                                                                & 0.004               & 0.004               & 30.74                                                 & 0.92    & 0.057 & $\lesssim$0.060        & NGC 1536                                                                         & \citet{Turatto98}, \citet{Zampieri03}        \\
& & & & & & & & & & \citet{Spiro14}\\
SN1999br    & 0.00323  & 0.002                                                                & 0.001               & 0.001               & 30.97                                                 & 0.83      & 0.063 & 0.000         & NGC 4900                                                                        & \citet{Hamuy03}, \citet{Pastorello04}, \\
& & & & & & &  & & & \citet{Gutierrez17a}\\
SN2002gd    & 0.00892  & <0.003                                                               & -                   & -                   & 32.87                                                 & 0.35     & 0.178 & 0.000       & NGC 7537                                                                         & \citet{Spiro14}, \citet{Gutierrez17a}        \\
SN2002gw     & 0.01028  & 0.012                                                                & 0.004               & 0.003               & 32.98                                                 & 0.23     & 0.051 & 0.000       & NGC 922                                                                         & \citet{Anderson14}, \citet{Galbany16}, \\
& & & & & & & & & & \citet{Gutierrez17a}\\
SN2003B     & 0.00424  & 0.017                                                                & 0.009               & 0.006               & 31.11                                                 & 0.28     & 0.072 & 0.180       & NGC 1097                                                                        & \citet{Blondin06}, \citet{Anderson14},\\
& & & & & & & & & & \citet{Galbany16}, \citet{Gutierrez17a}\\
SN2003fb    & 0.01754  & >0.017                                                               & -                   & -                   & 34.43                                                 & 0.12     & 0.482 & -       & UGC 11522                                                                        & \citet{Papenkova03}, \citet{Anderson14},\\
& & & & & & & & & & \citet{Gutierrez17a}\\
SN2003Z     & 0.0043   & 0.005                                                                & 0.003               & 0.003               & 31.70                                                 & 0.60      & 0.104 & 0.000       & NGC 2742                                                                         & \citet{Utrobin07}, \citet{Spiro14}        \\
SN2004fx    & 0.00892  & 0.014                                                                & 0.006               & 0.004               & 32.82                                                 & 0.24      & 0.274 & 0.000      & MCG -02-14-003                                                                   & \citet{Park04}, \citet{Anderson14},\\
& & & & & & & & & & \citet{Gutierrez17a}\\
SN2005cs    & 0.002    & 0.006                                                                & 0.003               & 0.003               & 29.46                                                 & 0.60      & 0.095 & 0.171       & M 51                                                                              & \citet{Pastorello06}, \citet{Pastorello09},\\
& & & & & & & & & & \citet{Spiro14}\\
SN2007aa    & 0.004887    & 0.032                                                                & 0.009               & 0.009               & 31.95                                                 & 0.27     & 0.070 & 0.000        & NGC 4030                                                                              & \citet{Anderson14}, \citet{Gutierrez17a},\\
& & & & & & & & & & This Work\\
SN2008bk    & 0.000767 & 0.007                                                                & 0.001              & 0.001               & 27.68                                                 & 0.13     & 0.052 & 0.000       & NGC 7793                                                                          & \citet{Vandyk12b}, \citet{Anderson14},\\
& & & & & & & & & & \citet{Spiro14} , \citet{Gutierrez17a}\\
SN2008in    & 0.005224 & 0.012                                                                & 0.005               & 0.005               & 30.60                                                 & 0.20   & 0.060 & 0.080          & NGC 4303                                                                         & \citet{Roy11}, \citet{Anderson14},\\ & & & & & & & & & & \citet{Gutierrez17a}          \\
SN2009N     & 0.003456 & 0.020                                                                 & 0.004               & 0.004               & 31.67                                                 & 0.11     & 0.056 & 0.100       & NGC 4487                                                                        & \citet{Takats14}, \citet{Anderson14},\\ & & & & & & & & & & \citet{Spiro14}, \citet{Gutierrez17a}      \\
SN2010id    & 0.01648  & -                                                                    & -                   & -                   & 32.86                                                 & 0.50      & 0.162 & 0.167       & NGC 7483                                                                        & \citet{GalYam11}, \citet{Spiro14}        \\
SN2012A     & 0.0025  & 0.011                                                                 & 0.004               & 0.004               & 29.96                                                 & 0.15    & 0.085 & $\sim$0.010        & NGC 3239                 
                                          & \citet{Tomasella13}, \citetalias{Jerkstrand15a}    \\
SN2012aw    & 0.0026  & 0.060                                                                 & 0.010               & 0.010               & 29.97                                                 & 0.03   & 0.074 & 0.143     & NGC 3351                  
                                          & \citet{Fraser12}, \citet{Bose13}, \citetalias{Jerkstrand14}, \citetalias{Jerkstrand15a}    \\
SN2012ec    & 0.00469  & 0.040                                                                 & 0.015               & 0.015               & 31.19                                                 & 0.13      & 0.071 & 0.372     & NGC 1084                                                                        & \citet{Barbarino15}, \citetalias{Jerkstrand15a}    \\
SN2013am    & 0.002692 & 0.015                                                                & 0.006               & 0.011               & 30.54                                                 & 0.40      & 0.066 & 1.705       & NGC 3623                                                                         & \citet{Zhang14, Tomasella18}        \\
SN2016aqf   & 0.004016    & 0.008                                                                    & 0.002               & 0.002               & 30.16                                                 & 0.27     & 0.146 &       $\lesssim$0.096                      &              NGC 2101                             & This Work              \\
SN2016bkv   & 0.002    & 0.0216                                                               & 0.0014              & 0.0014              & 30.79                                                 & 0.05     &  0.045 &$\lesssim$0.016       & NGC 3184                                                                        & \citet{Nakaoka18}, \citet{Hosseinzadeh18} \\

\hline
\end{tabular}
%\caption{SN II sample, with their respective host galaxies, used in this work. The redshifts are as reported in the references cited in column Ref. (directly measured from spectra). The M$_{\rm Ni}$ and $\mu$ values (together with the uncertainties) were obtained from the references cited in column Ref.. The v$_{\rm vir}$ and v$_{\rm 3k}$ values were obtained from http://leda.univ-lyon1.fr/ (these could differ from the values used in the references as this website is updated when more accurate measurements are available or because they change the adopted cosmology [H$_0$]).}
\label{tab:sn_sample}
\end{table*}

%% file: spectra_photospheric.tex
\begin{table*}
\centering
\caption{pEW for several lines during the optically thick phase and H$_{\alpha}$ FWHM. These values are not corrected for instrument resolution. Phase with respect to the explosion epoch.}
\tiny
\setlength\tabcolsep{3.5pt} % default value: 6pt
%\begin{adjustbox}{angle=90}
\begin{tabular}{lccccccccc}
\hline
Phase                & pEW(H$_{\beta}$) & pEW(Fe II 4924) & pEW(Fe II 5018) & pEW(Fe II 5169) & pEW(Na I D)    & pEW(Ba II 6142) & pEW(Sc II 6247) & pEW(H$_{\alpha}$) & FWHM(H$_{\alpha}$) \\
\multicolumn{1}{c}{} & [\AA]            & [\AA]           & [\AA]           & [\AA]           & [\AA]          & [\AA]           & [\AA]           & [\AA]             & [\AA]              \\ \hline
%5                    & -                & -               & -               & -               & -              & -               & -               & -                 & -                  \\
%6                    & -                & -               & -               & -               & -              & -               & -               & -                 & -                  \\
%8                    & -                & -               & -               & -               & -              & -               & -               & -                 & -                  \\
13                   & 31.7 $\pm$ 3.1   & -               & -               & -               & -              & -               & -               & 19.0 $\pm$ 0.9    & 189.7 $\pm$ 2.5    \\
16                   & 34.0 $\pm$ 2.0   & -               & 1.3 $\pm$ 0.1   & 12.7 $\pm$ 0.4  & -              & -               & -               & 29.2 $\pm$ 3.2    & 170.0 $\pm$ 2.6    \\
17                   & 33.9 $\pm$ 0.8   & -               & -               & 13.7 $\pm$ 0.4  & -              & -               & -               & 31.3 $\pm$ 3.1    & 181.3 $\pm$ 3.5    \\
22                   & 51.0 $\pm$ 2.6   & -               & 16.3 $\pm$ 0.6  & 19.3 $\pm$ 1.5  & -              & -               & -               & 46.0 $\pm$ 2.0    & 156.0 $\pm$ 1.7    \\
29                   & 32.3 $\pm$ 1.5   & -               & -               & -               & -              & -               & -               & 62.0 $\pm$ 7.0    & 140.0 $\pm$ 5.2    \\
37                   & 37.7 $\pm$ 1.2   & 6.3 $\pm$ 0.8   & 16.2 $\pm$ 0.8  & 23.7 $\pm$ 7.6  & 6.9 $\pm$ 1.3  & 3.2 $\pm$ 0.5   & 4.1 $\pm$ 1.0   & 62.3 $\pm$ 4.5    & 113.0 $\pm$ 6.0    \\
44                   & 43.3 $\pm$ 1.5   & 8.2 $\pm$ 0.4   & 19.3 $\pm$ 1.5  & 31.3 $\pm$ 2.3  & 9.4 $\pm$ 0.9  & 5.1 $\pm$ 0.4   & 3.9 $\pm$ 0.7   & 65.7 $\pm$ 3.2    & 100.3 $\pm$ 6.1     \\
51                   & 42.0 $\pm$ 1.0   & 11.7 $\pm$ 1.1  & 20.7 $\pm$ 1.2  & 31.0 $\pm$ 2.0  & 16.8 $\pm$ 0.3 & 6.0 $\pm$ 0.6   & 5.0 $\pm$ 0.5   & 65.3 $\pm$ 3.5    & 101.3 $\pm$ 3.8     \\
53                   & 47.3 $\pm$ 2.3   & 12.8 $\pm$ 0.7  & 22.0 $\pm$ 1.0  & 34.0 $\pm$ 1.7  & 19.7 $\pm$ 2.3 & 9.6 $\pm$ 1.0   & 7.8 $\pm$ 1.5   & 64.0 $\pm$ 2.6    & 93.7 $\pm$ 3.2     \\
54                   & 55.0 $\pm$ 3.6   & 11.7 $\pm$ 0.5  & 20.3 $\pm$ 0.6  & 33.7 $\pm$ 2.1  & 24.0 $\pm$ 1.7 & 8.9 $\pm$ 1.6   & 5.0 $\pm$ 0.4   & 68.7 $\pm$ 1.5    & 94.0 $\pm$ 3.0     \\
60                   & 45.3 $\pm$ 2.9   & 14.3 $\pm$ 0.5  & 23.0 $\pm$ 1.0  & 37.0 $\pm$ 3.0  & 24.0 $\pm$ 2.6 & 11.0 $\pm$ 1.1  & 7.1 $\pm$ 0.3   & 66.0 $\pm$ 2.6    & 88.3 $\pm$ 2.5     \\
71                   & 37.0 $\pm$ 1.7   & 17.7 $\pm$ 0.6  & 25.0 $\pm$ 1.0  & 40.3 $\pm$ 1.5  & 30.7 $\pm$ 2.1 & 15.3 $\pm$ 1.2  & 7.2 $\pm$ 0.5   & 62.0 $\pm$ 2.6    & 82.7 $\pm$ 5.0     \\ \hline
\end{tabular}
%\end{adjustbox}
\label{tab:spectra_photospheric}
\end{table*}

%% file: spectra_nebular.tex
\begin{table*}
\centering
\caption{FWHM for lines during the optically-thin phase.  Values are corrected for the instrument resolution.}
\setlength\tabcolsep{3.5pt} % default value: 6pt
\begin{tabular}{lccccc}
\hline
Phase                & FWHM([O I] 6300) & FWHM([O I] 6364) & FWHM(H$_{\alpha}$) & FWHM(He I 7065) & FWHM([Fe II] 7155) \\
\multicolumn{1}{c}{}[days]  & [\AA]             & [\AA]             & [\AA]              & [\AA]           & [\AA]              \\ \hline
155                  & -                 & -                 & 43.8 $\pm$ 0.6     & 47.8 $\pm$ 3.1  & 41.2 $\pm$ 2.1     \\
168                  & 29.5 $\pm$ 2.1    & 18.7 $\pm$ 2.1    & 40.0 $\pm$ 0.6     & 36.3 $\pm$ 2.6  & 36.3 $\pm$ 2.0     \\
202                  & 33.6 $\pm$ 1.5    & 20.8 $\pm$ 2.1    & 40.5 $\pm$ 0.6     & 30.3 $\pm$ 1.0  & 34.3 $\pm$ 1.2     \\
220                  & 38.2 $\pm$ 6.7    & 23.6 $\pm$ 1.2    & 38.2 $\pm$ 0.6     & 32.7 $\pm$ 1.0  & 34.7 $\pm$ 1.5     \\
259                  & 24.9 $\pm$ 2.5    & 17.8 $\pm$ 1.2    & 37.8 $\pm$ 0.6     & 24.4 $\pm$ 1.5  & 34.3 $\pm$ 1.5     \\
271                  & 24.4 $\pm$ 2.5    & 20.8 $\pm$ 2.1    & 39.4 $\pm$ 0.6     & 23.0 $\pm$ 1.5  & 34.7 $\pm$ 1.5     \\
285                  & 27.5 $\pm$ 2.1    & 20.8 $\pm$ 3.8    & 35.9 $\pm$ 0.6     & 28.7 $\pm$ 1.5  & 32.4 $\pm$ 1.2     \\
303                  & 28.2 $\pm$ 2.5    & 22.2 $\pm$ 4.0    & 35.4 $\pm$ 0.6     & 27.0 $\pm$ 0.6  & 29.5 $\pm$ 1.5     \\
330                  & 28.2 $\pm$ 1.5    & 24.9 $\pm$ 2.1    & 35.9 $\pm$ 0.6     & 28.2 $\pm$ 2.1  & 34.7 $\pm$ 1.2     \\
351                  & 39.7 $\pm$ 3.6    & 19.8 $\pm$ 4.4    & 37.8 $\pm$ 0.6     & 70.9 $\pm$ 9.0  & 43.4 $\pm$ 4.0     \\ \hline
\end{tabular}
\label{tab:spectra_nebular}
\end{table*}

%% file: nife_ratio.tex
\begin{table}
\caption{Ni/Fe abundance ratio values used in this work.}
\centering
\begin{tabular}{lccc}
\hline
SN        & Ni/Fe & $\sigma_-$ & $\sigma_+$ \\ 
\hline \hline
SN1997D   & 0.079 & 0.025      & 0.014      \\
SN2003B   & 0.057 & 0.021      & 0.018      \\
SN2005cs  & 0.084 & 0.012      & 0.012      \\
SN2007aa  & 0.074 & 0.006      & 0.006      \\
SN2008bk  & 0.046 & 0.042      & 0.017      \\
SN2009N   & 0.101 & 0.018      & 0.017      \\
SN2012A   & 0.028 & 0.022      & 0.016      \\
SN2012aw  & 0.084 & 0.022      & 0.016      \\
SN2012ec  & 0.2   & 0.07       & 0.07       \\
SN2013am  & 0.108 & 0.017      & 0.018      \\
SN2016aqf & 0.081 & 0.010      & 0.009      \\ \hline
\end{tabular}
\label{tab:nife}
\end{table}